\documentclass[a4paper,useAMS,usenatbib]{mnras}

\usepackage{graphicx}
\usepackage{setspace}
\usepackage{natbib}
\usepackage{color}
\usepackage{amssymb,amsmath}
\usepackage{times}
\usepackage{hyperref}

\def\nmem{109}
\def\phione{\phi_1}
\def\phitwo{\phi_2}

\def\muone{\mu_{\phione}}
\def\mutwo{\mu_{\phitwo}}

\newcommand{\Gaia}{{\it Gaia}}
\newcommand{\gaia}{\textit{Gaia} }

\definecolor{darkred}{rgb}{0.64, 0.0, 0.0}
\definecolor{darkpurple}{rgb}{0.6, 0.0, 0.6}

\title[All-sky view of the Orphan Stream]{Piercing the Milky
  Way: an all-sky view of the Orphan Stream}

\author[Koposov et al.]{
\parbox{\textwidth}{
\Large
S.~E.~Koposov$^{1,2}$,
V.~Belokurov$^{2,3}$,
T.~S.~Li$^{4,5}$,
C.~Mateu$^{6}$,
D.~Erkal$^{7}$,
C.~J.~Grillmair$^{8}$,
D.~Hendel$^{9}$,
A.~M.~Price-Whelan$^{10}$,
C.~F.~P.~Laporte$^{11}$,
K.~Hawkins$^{12}$,
S.~T.~Sohn$^{13}$,
A.~del~Pino$^{13}$,
N.~W. ~Evans$^{2}$,
C.~T.~Slater$^{14}$,
N.~Kallivayalil$^{15}$,
J.~F.~Navarro$^{16}$
\begin{center} (The OATs: Orphan Aspen Treasury Collaboration) \end{center}
}
\vspace{0.4cm}
\\
\parbox{\textwidth}{
$^{1}$ Carnegie Mellon University, 5000 Forbes Ave, Pittsburgh, PA, 15213, USA\\
$^{2}$ Institute of Astronomy, University of Cambridge, Madingley Road, Cambridge CB3 0HA, UK\\
$^{3}$ Center for Computational Astrophysics, Flatiron Institute, 162 5th Avenue, New York, NY 10010, USA\\
$^{4}$ Fermi National Accelerator Laboratory, P.O.\ Box 500, Batavia, IL 60510, USA\\
$^{5}$ Kavli Institute for Cosmological Physics, University of Chicago, Chicago, IL 60637, USA\\
$^{6}$ Departamento de Astronom\'{i}a, Instituto de F\'{i}sica, Universidad de la Rep\'{u}blica, Igu\'a 4225, CP 11400 Montevideo, Uruguay\\
$^{7}$ Department of Physics, University of Surrey, Guildford GU2 7XH, UK\\
$^{8}$ IPAC, Mail Code 314-6, Caltech, 1200 E. California Blvd., Pasadena, CA 91125, USA\\
$^{9}$ Department of Astronomy and Astrophysics, University of Toronto, 50 St. George Street, Toronto, Ontario, M5S 3H4, Canada\\
$^{10}$ Department of Astrophysical Sciences, Princeton University, 4 Ivy Lane, Princeton, NJ 08544, USA\\
$^{11}$ CITA National Fellow, Department of Physics and Astornomy, University of Victoria, 3800 Finnerty Road, Victoria, BC, V8P 5C2, Canada\\
$^{12}$ Department of Astronomy, The University of Texas at Austin, 2515 Speedway Boulevard, Austin, TX 78712, USA\\
$^{13}$ Space Telescope Science Institute, 3700 San Martin Drive, Baltimore 21218, USA\\
$^{14}$ Department of Astronomy, University of Washington, Box 351580, Seattle, WA 98195, USA\\
$^{15}$ Department of Astronomy, University of Virginia, 530 McCormick Road, Charlottesville, VA 22904, USA\\
$^{16}$ Senior CIfAR Fellow, Department of Physics and Astronomy, University of Victoria, 3800 Finnerty Road, Victoria, BC, V8P 5C2 Canada\\
}
}

\begin{document}

\maketitle

\label{firstpage}

\begin{abstract}
We use astrometry, broad-band photometry and variability information
from the Data Release 2 of ESA's \gaia mission (GDR2) to identify
members of the Orphan Stream (OS) across the whole sky. The stream is
traced above and below the celestial equator and in both Galactic
hemispheres, thus increasing its visible length to $\sim210^{\circ}$
equivalent to $\sim150$ kpc in physical extent. Taking advantage of
the large number of RR Lyrae stars in the OS, we extract accurate
distances and proper motions across the entire stretch of the tidal
debris studied. As delineated by the GDR2 RR Lyrae, the stream
exhibits two prominent twists in its shape on the sky which are
accompanied by changes in the tangential motion. We complement the RR
Lyrae maps with those created using GDR2 Red Giants and the DECam
Legacy Survey Main Sequence Turn-Off stars. The behavior of the OS
track on the sky is consistent across all three tracers employed. We
detect a strong non-zero motion in the across-stream direction for a
substantial portion of the stream. Such a misalignment between the
debris track and the streaming velocity cannot be reproduced in a
static gravitational potential and signals an interaction with a
massive perturber.

\end{abstract}


\section{Introduction}


%
\begin{figure*}
  \centering
  \includegraphics[width=0.95\textwidth]{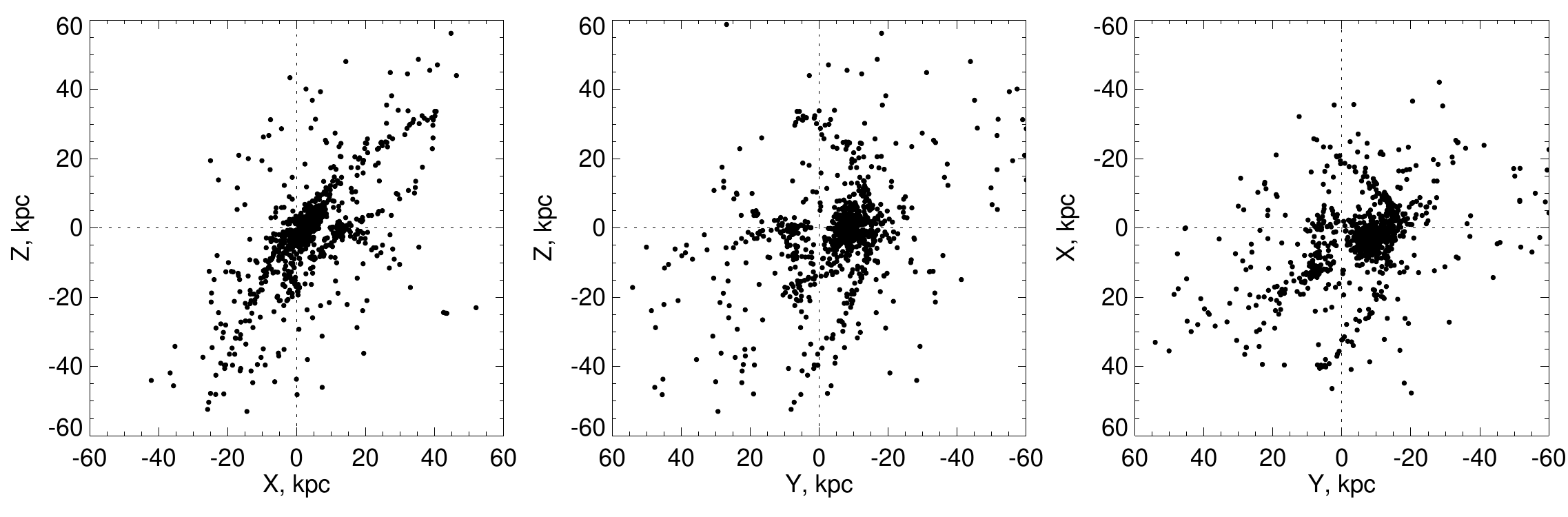}
  \caption[]{Galactocentric distributions of the $\sim1,000$ \gaia
    DR2 RR Lyrae stars projected close to the OS on the sky. Only
    stars with $|\phi_2|<4^{\circ}$ and $-300<\frac{v_l}{1\,{\rm km\,s^{-1}}} <-100$, $-300<\frac{v_b}{1\,{\rm km\,s^{-1}}}
    <300$ are shown. {\it Left:} $X,Z$ plane. {\it Middle:} $Y,Z$
    plane. {\it Right:} $X,Y$ plane. Note a prominent, narrow and long
    stream-like over-density visible in all three panels.}
   \label{fig:xyzgdr2}
\end{figure*}

Long and narrow streams made up of stars tidally removed from low-mass
satellites offer a direct, powerful and versatile means of probing the
Galaxy's gravity field \citep[see e.g.][]{Donald1982, Kuhn1993,
  Donald1995,Johnston1996,Helmi1999,Johnston1999,Murali1999}. Discussed
predominantly from a theoretical standpoint in the last decades of the
past century, stellar streams started to be discovered in large
numbers relatively recently
\citep[e.g.][]{Ibata2001,Odenkirchen2001,Newberg2002,
  Majewski2003,FOS,Grillmair2006}. Deep wide-area imaging data played
a crucial role: the Sloan Digital Sky Survey \citep[SDSS,
  see][]{Gunn1998,York2000,SDSS_DR8,SDSS_DR12} opened the floodgate
\citep[e.g.][]{OS_C,OS_V,Grillmair2009,Newberg2009,Koposov2012,
  Bonaca2012} and other surveys followed
\citep[e.g.][]{Koposov2014,Bernard2016,Balbinot2016}, culminating most
recently with the discovery of 11 new streams in the Dark Energy
Survey data \citep[DES, see][]{DES2005,DES2016} by \citet{Shipp2018}.

The arrival of the \gaia data \citep[see][]{Prusti2016, Brown2018} has
helped to push the stellar stream identification into even lower
surface brightness regime through the addition of all-sky
high-precision astrometry
\citep[see][]{Myeong2018,Ibata2018,Koppelman2018,Malhan2018,Adrian2018,CloudsArms}. 
 \Gaia's
parallaxes and proper motions are valuable not only because they can
be used to beat down the overpowering intervening Galactic
foreground. The utility of a stellar tidal stream springs as soon as
the kinematic information is added as demonstrated convincingly by
\citet{Koposov2010}. For example, the streaming velocity, i.e. the
component of the motion in the direction aligned with the stream, sets
the normalization for the mass inside of the debris orbit, while the
across-stream velocity (i.e. the velocity component tangential to the
stream path) informs of possible perturbations \citep[see
  e.g.][]{Erkal2018}.

One of the first stellar streams discovered as the SDSS unleashed its
data on the unsuspecting community was Orphan
\citep[see][]{OS_C,OS_V}. Ten years later, the stream remains so,
literally, as no plausible progenitor has been identified despite
numerous attempts
\citep[e.g.][]{Fellhauer2007,Jin2007,Sales2008,Casey2013,Casey2014,Grillmair2015}.
While the stream's known length has been extended somewhat below the
celestial equator, no trace of it has so far been found in the
Southern Galactic hemisphere. After some initial confusion, the orbit
of the stream was clarified in \citet{Newberg2010}, who took advantage
of the wide-area, low-resolution spectroscopy provided by the SDSS and
the Sloan Extension for Galactic Understanding and Evolution
(SEGUE). An equally important role was played by the Blue Horizontal
Branch (BHB) stars used by \citet{Newberg2010} to nail down the
distances to the debris and thus reveal the 3-D shape of the
stream. Given the distance gradient and the pattern of line-of-sight
velocities, the direction of the stream motion became immediately
apparent. In Galactic coordinates, the OS is moving up (as defined by
the Galactic North), hence the part of the stream observed by the SDSS
is leading the presumed missing section below the Milky Way's disc. By
fitting orbits to the detected portions of the OS, \citet{Newberg2010}
determined the stream's extrema, placing the peri-center at $\sim16$
kpc and the apo-center at $\sim90$ kpc.

Conveniently, not only does the OS contain plenty of BHBs but it also
boasts a prominent RR Lyrae (RRL) population \cite[see
  e.g.][]{Sesar2013}. In the optical, RR Lyrae are slightly better
standard candles than BHB stars \citep[see][]{Fermani2013,Sesar2017}. Importantly, however, RR Lyrae suffer
little contamination due to their unique light curves, while BHBs can
be confused with Blue Stragglers and occasionally with White Dwarfs
and/or QSOs \citep[see e.g.][]{Deason2014}. An additional perk of
using RR Lyrae is that a rough estimate of the star's metallicity can
be obtained based on the light curve shape. \citet{Sesar2013}
identified $\sim30$ RR Lyrae likely belonging to the OS using a
combination of Catalina Real-Time Sky Survey (CRTS), Lincoln Near
Earth Asteroid Research (LINEAR) survey, and Palomar Transient Factory
(PTF) data. While the RRLs are scattered rather sparsely along the
stream, approximately one member star per 2-3 degrees on the sky, they
are nonetheless easily discernible from the bulk of the Milky Way
stellar halo. The RR Lyrae's period-luminosity relation is
exceptionally tight and steep in the near-infrared. Accordingly,
\citet{Hendel2018} used the Spitzer IRAC 3.6$\mu$m to follow-up the
RRLs identified in \citet{Sesar2013} and measure distances to the
individual OS stars with uncertainty of order $\sim2.5\%$.

\begin{figure*}
  \centering
  \includegraphics[width=0.9\textwidth]{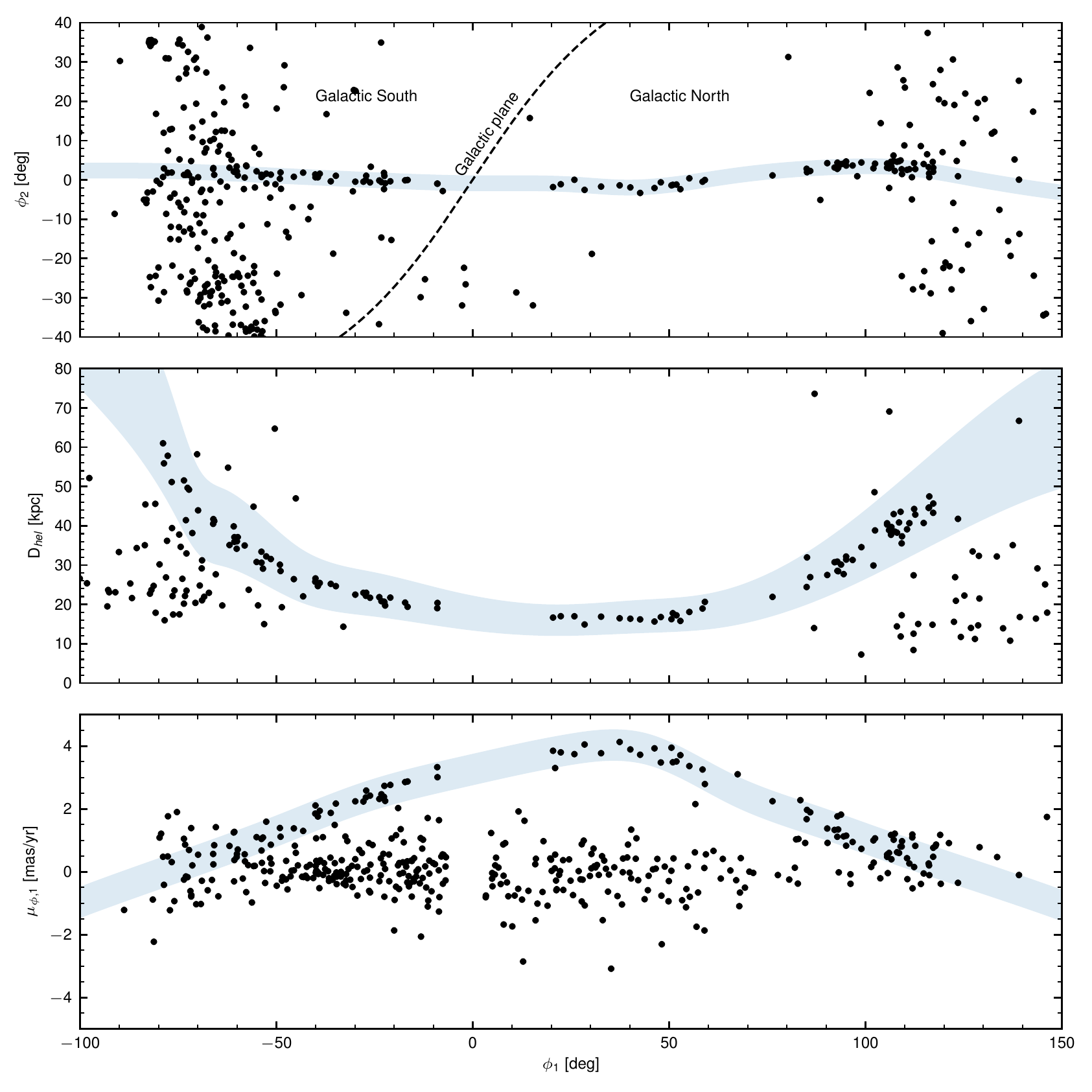}
  \caption[]{The selection of Orphan Stream RR Lyrae in various
    projections of the data. Each of the panels shows the sample
    of stars based on the selection of stars from the other panels. The common
	selection criteria applied to all the
	panels are the astrometric excess noise, parallax, and BP/RP excess cuts described
	in the text as well as selection based on proper motion across the stream $-0.5\,{\rm mas\,yr}^{-1}<\mutwo<0.9\,\,{\rm mas\,yr}^{-1}$.
     {\it
      Top panel:} The distribution of RR Lyrae on the sky in stream
    coordinates. The star selection used for this plot is based on the
    distance and proper motion information. The transparent blue band
    shows our selection region for the stream track on the sky.
    {\it Middle panel:} The
    heliocentric distance to RR Lyrae as a function of angle along the
    stream. The sample of stars uses the selection based on the track
    on the sky and proper motion (as defined in the bottom and top
    panels). The blue band shows the selection region for distances.
    {\it Bottom panel:} The proper motion along $\phione$
    (with the correction for the solar motion applied) vs the angle
    along the stream. The sample of stars shown on the panel is based on
    selections shown on top and middle panels. The blue band shows the
    proper motion selection region.  }
   \label{fig:selection}
\end{figure*}
\begin{figure*}
  \centering
  \includegraphics[width=0.97\textwidth]{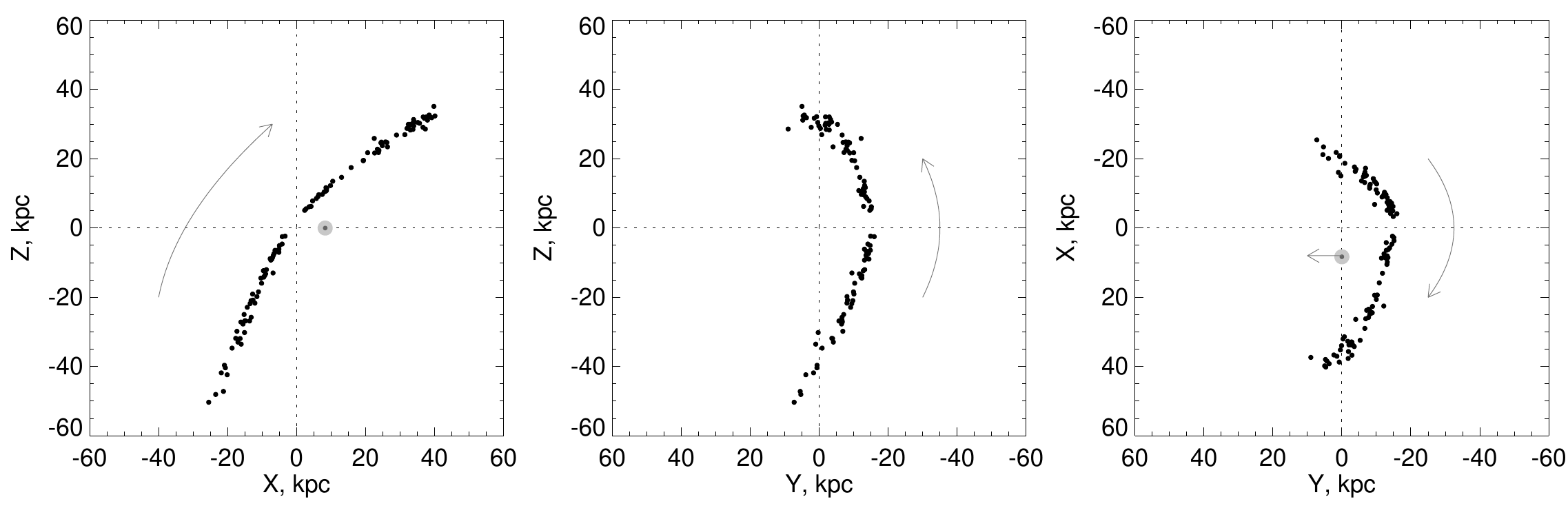}
  \caption[]{Same as Figure~\ref{fig:xyzgdr2} but limited to likely OS
    member RR Lyrae (see Figure~\ref{fig:selection} and main text for
    details). Long curved arrow shows the direction of motion of the
    OS, while the short straight arrow indicates the motion of the Sun
    (shown as grey filled circle at $(X,Y,Z)=(8.3,0,0)$.}
   \label{fig:xyzmem}
\end{figure*}
\begin{figure*}
  \centering
  \includegraphics[width=0.9\textwidth]{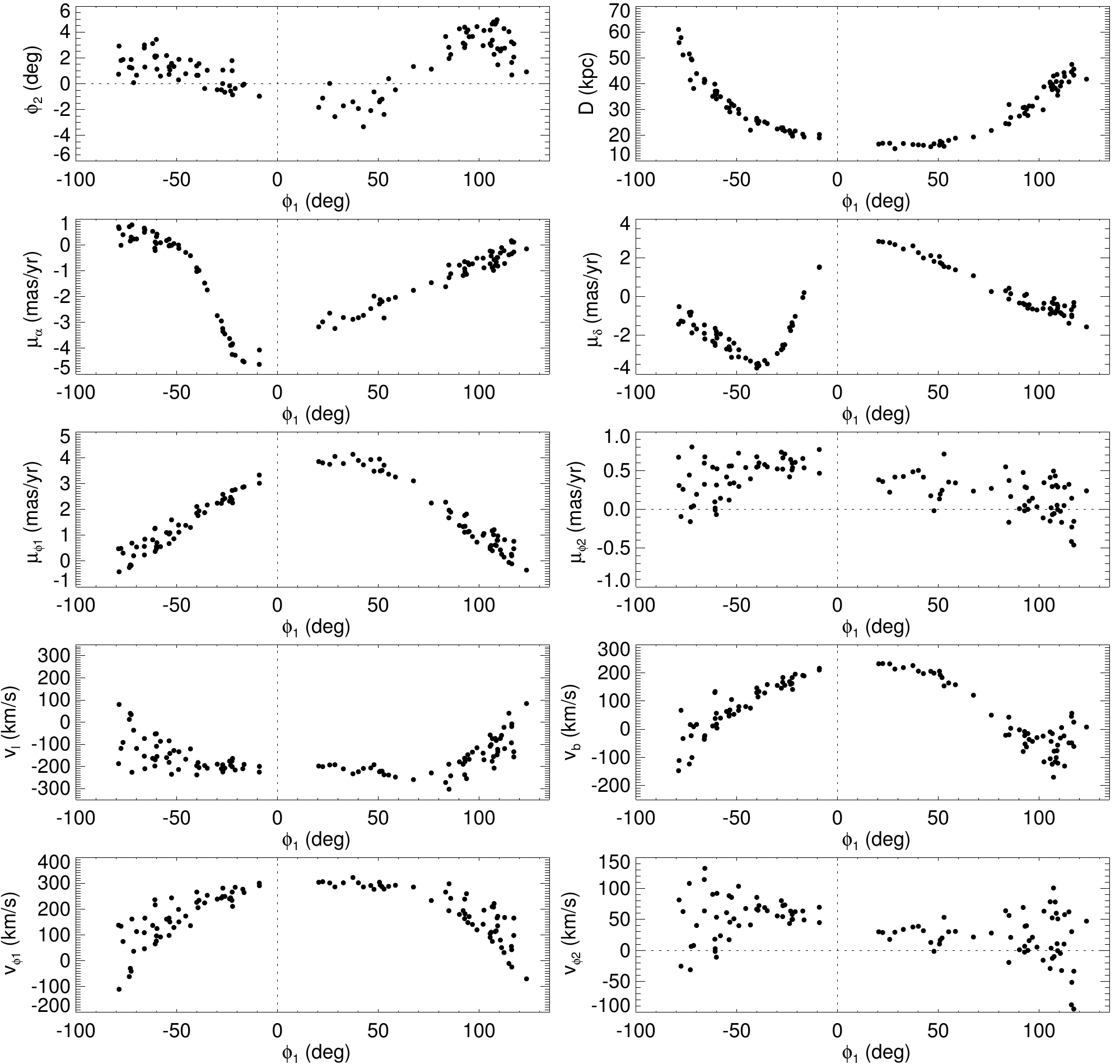}
  \caption[]{Phase-space projections of the likely OS member RR Lyrae
    as a function of the stream longitude $\phi_1$. {\it Top left:}
    Positions of the OS stars on the sky in the stream-aligned
    coordinates $\phi_1, \phi_2$. {\it Top right:} Heliocentric
    distance as a function of $\phi_1$. {\it Second row, left:} Right
    ascension component of proper motion $\mu_{\alpha}$. {\it Second
      row, right:} Declination component of proper motion
    $\mu_{\delta}$. {\it Third row, left:} Proper motion along the
    stream (strictly speaking, along the longitude axis of the
    stream-aligned coordinates) $\mu_{\phi,1}$. {\it Third row,
      right:} Proper motion across the stream $\mu_{\phi, 2}$ {\it
      Fourth row, left:} Velocity component along Galactic longitude
    $v_l$. {\it Fourth row, right:} Velocity component along Galactic
    latitude $v_b$. {\it Bottom left:} Along-stream velocity
    $v_{\phi,1}$. {\it Bottom right:} Across-stream velocity
    $v_{\phi,2}$. Note that all proper motions are corrected for the
    effect of the Solar reflex motion.}
   \label{fig:memother}
\end{figure*}

In this paper we provide the first all-sky map of the Orphan Stream
constructed with a range of stellar tracers. We start with the \gaia
DR2 (GDR2) RR Lyrae, the largest set of RR Lyrae available to date
\citep[see][]{Clementini2018,Holl2018}. The \gaia DR2 RR Lyrae are
identified and characterized in the optical where the
period-luminosity relation is rather flat, thus yielding distance
estimates inferior to those obtained using infrared data. Nonetheless
they are very competitive and superior to most other tracers, with
typical uncertainties $<10\%$. Despite the high purity (outside of the
Galactic plane) and impressive completeness (for heliocentric
distances below 70 kpc), the GDR2 RR Lyrae sample remains relatively
unexplored in terms of Galactic halo studies \citep[e.g.][]{Iorio2018,
  Torrealba2018}. To complement the OS RR Lyrae detections we also
study the stream with other stellar tracers, including GDR2 Red Giants
and Main Sequence Turn-off stars in Dark Energy Survey
\citep{DESDR1_2018} and DECam Legacy Survey
\citep[DECaLS, ][]{DECALS2018} data.

This Paper is organised as follows. Section~\ref{sec:rrl} presents the
details of the RR Lyrae selection from the \gaia DR2 data as well as
the selection of the likely Orphan Stream members. That Section also
gives the distributions of the OS RR Lyrae in various projections of
the phase-space spanned by the three spatial coordinates and two
tangential velocities. Section~\ref{sec:rgb} maps the stream using
\gaia Red Giant stars as well as using the matched filter technique
applied to the deep DECaLS imaging. Additionally, the Section shows
the OS Color-Magnitude Diagram and discusses the subset of the stream
members with available SDSS
spectroscopy. Finally. Section~\ref{sec:disc} puts the new detections
of the stream into context. Concluding remarks are provided in
Section~\ref{sec:conc}.

\section{The Orphan Stream with \gaia DR2 RR Lyrae}
\label{sec:rrl}

We select a high-purity all-sky sample of RR Lyrae candidate stars
from two separate source catalogues released as part of the \gaia DR2
\citep[][]{Prusti2016, Brown2018}. More precisely, tables
\texttt{vari\_classifier\_result} and \texttt{vari\_rrlyrae}
\citep[see][]{Clementini2018,Holl2018} are joined, removing the
duplicates, and the stellar astrometry and photometry are obtained
from the main \texttt{gaia\_source} catalogue. We provide the query
for this RR Lyrae dataset in the Appendix~\ref{sec:rrl_query}. We have
culled potential interlopers requiring
\textit{phot\_bp\_rp\_excess\_factor} to be less than 1.5 and have
assumed $A_G/E(B-V)=2.27$ and $M_G=0.64$ for the extinction coefficient and
the RRL absolute magnitude in the \Gaia's $G$ band respectively
\citep[see][for further details]{Iorio2018}. We also required a
negligible parallax $\varpi < \textrm{max}(3\,\sigma_\varpi, 0.1)$ and a small
astrometric excess noise ($\textrm{AEN}<1$).  Additionally, we have assumed the
Sun's distance from the Galactic center of 8.3 kpc
\citep[][]{Gillessen2009}, the Sun's height above the Galactic mid-plane
of 27 pc \citep[][]{Chen2001}, the Local Standard of Rest (LSR) of 220 km
s$^{-1}$ \citep[][]{Bovy2015}, and the Sun's peculiar motion as given
in \citet{LSR}.

To select the likely OS members, we translate the stellar celestial
coordinates in equatorial system $(\alpha, \delta)$ into a new
coordinate system $(\phi_1, \phi_2)$ aligned with the stream
\citep[see e.g.][]{Koposov2010}. More specifically, based on the OS
detections reported in \citet{OS_V, OS_C, Newberg2010}, we use a great
circle with a pole at $(\alpha_{\rm OS}, \delta_{\rm OS})=(72\degr,
-14\degr)$. The origin of this stream's coordinate system is chosen to
be at $(\alpha_0, \delta_0)=(191\fdg10487, -62\fdg86084)$,
near the position where the equator of this coordinate system (and the OS) crosses the Galactic plane. We
provide the rotation matrix to this coordinate system in
Appendix~\ref{sec:rotation_matrix}.  Finally, $\phi_1$ increases in
the direction of the stream's motion, i.e. from the Galactic South to
the Galactic North. Several attempts at the OS kinematic
characterization can be found in \citet{Newberg2010} and
\citet{Sohn2016}. Given the OS's line-of-sight velocity and the
HST-based proper motions, we notice that the stream's velocity
component along Galactic longitude $v_l$ remains negative
irrespective of the debris position on the sky, i.e. the stream is in
prograde motion with respect to the Galactic rotation. Accordingly, we
explore the distribution of the possible OS members selected with the
following simple cuts.

\begin{equation}
  \begin{aligned}
    |\phi_2| &< 4^{\circ}\\
    -300<\frac{v_l}{1 {\rm km\, s^{-1}}} &<-100\\
    -300<\frac{v_b}{1 {\rm km\, s^{-1}}} &<300
  \end{aligned}
\end{equation}

Figure~\ref{fig:xyzgdr2} shows the distributions of $\sim$1,000 RRL
stars selected with the above cuts in Galactocentric coordinates
$X,Y$ and $Z$; here $X$ points to the Galactic anti-center and the Sun
is at $(X,Y,Z)=(8.3,0,0)$. A long and narrow arc of RRL stars is
visible in all three projections, crossing the Galaxy from the North
(where it is seen at $Z>0$ and $X>0$) to the South (where the signal
is at $Z<0$ and $X<0$). The stream spans a gigantic $\sim$150 kpc,
traveling in almost uninterrupted fashion through the Milky Way,
coming as close as $\sim$15 kpc to its center and reaching as far as
$\sim$50 kpc into the halo. Yet neither an obvious progenitor nor any
sign of the stream's apo-center is apparent. While the Northern
portion of the stream had been seen before with SDSS and Pan-STARRS1 \citep[PS1;][]{ps1}, the view
of its Southern Galactic section uncovered here with \gaia DR2 is
entirely new.

We can further clean the OS membership by looking closely at the
behavior of the RRL stars close to the equator of the stream's
coordinate system. To this end, Figure~\ref{fig:selection} shows the
distribution of RR Lyrae as a function of the along-stream coordinate
$\phione$. The top panel shows the distribution of RRL on the sky, the middle
one shows the distance as a function of the angle along the stream and the bottom one shows the proper motion of the stream stars along the $\phione$ direction (after correcting  for the solar reflex motion).
Each panel of Figure~\ref{fig:selection} shows a sample of stars
selected after applying the selection masks from the other two panels
(the masked areas are highlighted in pale blue in each panel), i.e. the stars for the top panel were selected based on their proper motion and distance. An additional proper motion selection was applied to all stars in this figure to limit $-0.5<\frac{\mutwo}{1\,{\rm mas\,yr^{-1}}}<0.9$. The masks are specified by first adopting reference stream tracks in proper motion, distance and on the sky.
The reference stream tracks are defined as natural cubic splines.
For the rest of the paper, we will refer to these spline tracks in distance, proper motion and on the sky as $\hat{D}(\phione)$,
$\hat{\mu}_{\phi,1}(\phi_1)$ and $\hat{\phi}_2(\phi_1)$
The positions of the spline knots and the corresponding values
were mostly chosen manually based on the RR Lyrae distribution seen on Figure~\ref{fig:selection} and are provided in
Tables ~\ref{tab:spline_track}, \ref{tab:dist_track},
\ref{tab:pm_track} in the Appendix.
The boundaries of the masks are chosen to wrap around the reference tracks of the stream and are defined as $0.75\,\hat{D}(\phione)<{D}<1.25\,\hat{D}(\phione)$, $|\muone-\hat{\mu}_1(\phione)|<1\,$ mas\,yr$^{-1}$, $|\phitwo-\hat{\phi}_2(\phione)|<1\degr$ for distance, proper motion and on-sky track respectively.
 Applying all three masks shown in the Figure
yields the total of \nmem\ OS candidate members. Table~\ref{tab:rr_list}
lists all RRL stars identified using the procedure described above
together with their positions and heliocentric distances.

As the top panel of the Figure illustrates, the stream spans
some $\sim210^{\circ}$ in $\phi_1$ while exhibiting a slight bend of
several degrees in $\phi_2$. The heliocentric distance of the Orphan
debris changes slowly within $-50^{\circ}<\phi_1<50^{\circ}$, but
beyond that, the stream appears to shoot rapidly to large distances,
reaching $\sim50$ kpc on either end. Finally, a clear kinematic
pattern is visible in the bottom panel of Figure~\ref{fig:selection},
where the stream $\mu_{\phi,1}$ proper motion arches up to $\sim 4$\,mas\,yr$^{-1}$ around $\phi_1\sim40^{\circ}$ from values close to $\sim0$
mas\,yr$^{-1}$ near the tips of the stream. Note that in all three
projections, the width of the stream also changes as a function of
$\phi_1$.
Some of this variation may be due to the intrinsic evolution
of the debris density along the stream \citep[see, e.g.,][]{stray}, but
much of it may plausibly be associated with the variations in the
uncertainties, especially for the proper motions.

Figure~\ref{fig:xyzmem} is a companion to Figure~\ref{fig:xyzgdr2} as
it also shows the Galactocentric positions of the \gaia RR Lyrae
stars, but now only those selected using the masks described
above. The leftmost panel of Figure~\ref{fig:xyzmem} emphasizes the high eccentricity of
the OS as the stream appears highly stretched in this
projection. Noticeable in the middle and right panels of this Figure is
the change in the stream's curvature between the Southern and the
North Galactic hemispheres. The view of the OS in other phase-space
projections can be found in Figure~\ref{fig:memother}.

The top left panel of Figure~\ref{fig:memother} presents the on-sky
positions of the likely OS members in the stream-aligned coordinate
system (note the large aspect ratio). We remark on the striking change of the
stream curvature around $20^{\circ}<\phi_1<50^{\circ}$ --- immediately
after the stream re-emerges after passing through the Galactic disc
--- where the debris behavior changes from gently sloping down to
sharply rising up in $\phi_2$. Additionally, in the Northern end of
the stream, at $\phi_1\sim100^{\circ}$ there appears to be a curious
hook downwards. There exists a corresponding change in the debris
kinematics. For example, proper motion components $\mu_{\alpha}$ and
$\mu_{\delta}$ also show a switch in the gradient as a function of
$\phi_1$ around $\phi_1\sim20^{\circ}$. Note, however, that the
evolution of the physical velocity components $v_l$ and $v_b$ (and
also $v_{\phi,1}$ , $v_{\phi,2}$, $\muone$ and $\mutwo$) is
considerably smoother. Therefore, most of the change in the proper
motions (shown in the second row from the top) is due to the change in
the line-of-sight distance and the contribution of the solar reflex
motion.

One extraordinary feature observed in Figure~\ref{fig:memother} is the
strong deviation of the across-stream velocity component $v_{\phi,2}$
when comparing the northern part of the stream
$50^{\circ}<\phi_1<100^{\circ}$ to the southern extension
$-50^{\circ}<\phi_1<0^{\circ}$. In the southern extension ---
corresponding to the locations under the Galactic plane ---
$v_{\phi,2}\sim50$ km s$^{-1}$, while in the North $v_{\phi,2}$ is
mostly consistent with 0 km s$^{-1}$. Note that in a static
gravitational potential, the stars in a cold stellar stream are
expected to move along the direction delineated by the stream track,
i.e. with $v_{\phi,2}\sim0$\,km\,s$^{-1}$ \citep[see
  e.g.][]{Koposov2010,Erkal2018}. Exceptions to this rule are small
regions around the location of the progenitor as well as near the
apo-center. We conjecture that the observed misalignment between the
motion of the RR Lyrae and the stream path is indicative of a strong
perturbation experienced by the stream in the not-so-distant past. A
large and massive deflector is required to divert orbits of stars in
such an extended portion of the stream.

\section{OS stars in GAIA, DECaLS and SDSS}
\label{sec:rgb}

\subsection{Track on the sky}

Guided by the RR Lyrae detections, we now chart the OS debris using
other stellar tracers, while taking advantage of the stream track
information in proper motion, distance and position on the sky. As a
first step we try to map the stream using the red giant branch stars
in the \gaia data.  We start by identifying the OS signal in the \gaia
color-magnitude space and obtain the background subtracted Hess
diagram of Orphan members.  We use \Gaia's $G_{\rm BP}$, $G_{\rm RP}$,
and $G$ photometry \citep[see][for additional details]{Evans2018}
corrected for extinction following the prescription of
\citet{Babusiaux2018}. As we have measured the distance evolution
along the stream from the RR Lyrae (see previous Section and
Table~\ref{tab:dist_track} for the definition of the distance spline
as a function of the angle along the stream), we correct all the
magnitudes of the stars by the the distance modulus expected at a
given $\phione$. The background subtracted Hess diagrams for the
stream are constructed by selecting stars within 0.7 degrees of the
stream track (as defined in Table \ref{tab:spline_track}) for the
northern stream and within 1 degree for the southern stream. The areas
located between 2 and 4 degrees away from the track were used as a
background for the Northern stream, and 2 and 6 degrees for the
Southern. Additionally, we apply the proper motion selection based on
splines presented in Figure~\ref{fig:selection} and
Table~\ref{tab:pm_track}:
$|\muone-\hat{\mu}_{\phi,1}(\phione)|<0.5$\,mas\,yr$^{-1}$ for both
streams and $-0.5<\frac{\mutwo}{1\, {\rm mas\,yr^{-1}}}<1$ and
$|\mutwo|<0.5$\,mas\,yr$^{-1}$ for the Southern and Northern streams
respectively.

\begin{figure}
\includegraphics{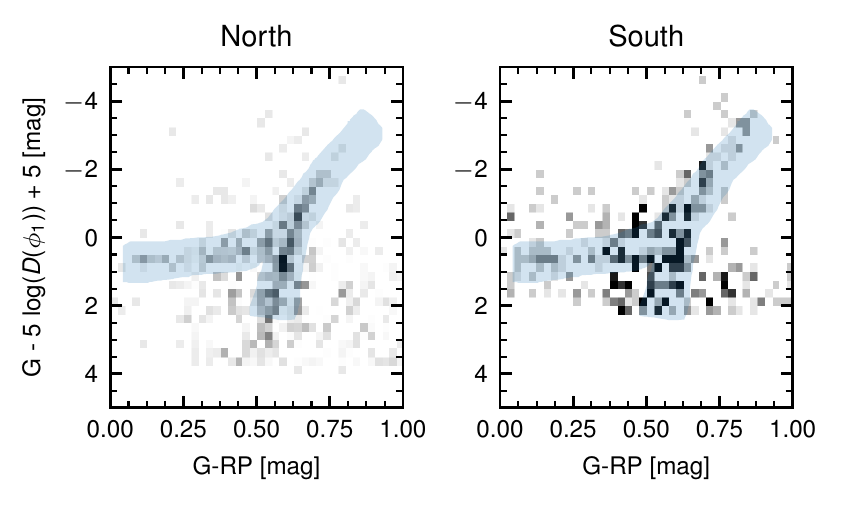}
\caption{The \gaia background-subtracted Hess diagrams of the
  Northern and Southern parts of the stream. The stars have been
  selected based on proximity to the stream (within 0.7 degrees of the
  stream for the North and 1 degree for the South) and proximity to
  the proper motion track
  $|\muone-\hat{\mu}_{\phi,1}(\phione)|<0.5$\,mas\,yr$^{-1}$. The region used to determine the background Hess
  was $2\degr<|\phi_2-\hat{\phitwo}(\phione)|<4\degr$ for
  the Northern stream and $2\degr<|\phi_2-\hat{\phitwo}(\phione)|<6\degr$ for the
  Southern. The magnitudes of all stars have been corrected by the
  distance modulus expected from the RR Lyrae distance track at a given $\phione$. The left panel includes the
  data with $30\degr<\phi_1<150\degr$. The right panel includes data from
  $-40\degr<\phi_1<-20\degr$.
  The blue shaded area shows the CMD mask that we adopt for further analysis of the \gaia data.
    }
\label{fig:cmd_gaia}
\end{figure}

The resulting CMD for both parts of the stream is shown in the
left panel of Figure~\ref{fig:cmd_gaia}. Focusing on the CMD of the Northern stream shown on the left panel, we observe an
unambiguous detection of a prominent red giant branch (RGB) and a
horizontal branch (HB) of the Orphan Stream. { This CMD signal can now
be used to identify the likely OS member stars more
efficiently as we do not expect the CMD of stream stars to vary significantly 
along the stream.} Accordingly, we draw the mask around the main CMD
features seen in the left panel of Figure~\ref{fig:cmd_gaia} and
select only stars lying within the mask. The mask is shown by a blue shaded region
in this Figure.

\begin{figure*}
  \centering \includegraphics[width=1\textwidth]{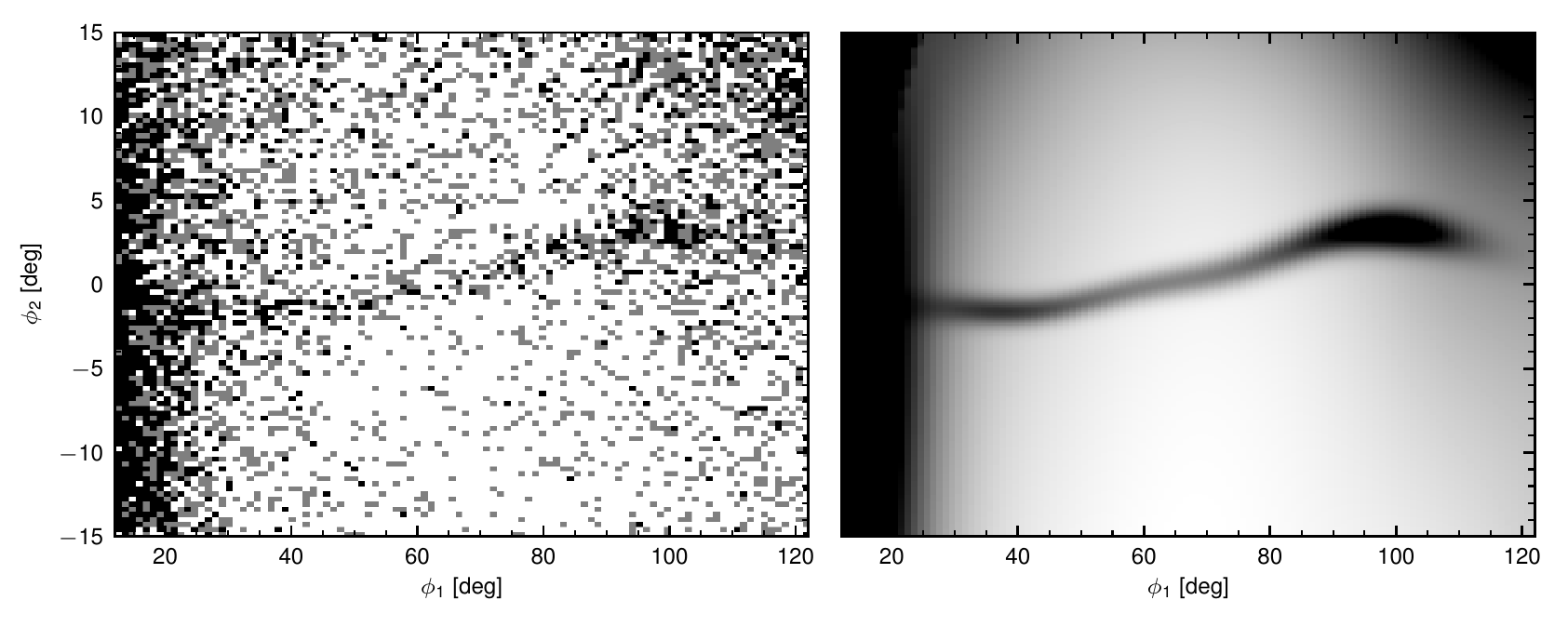}
  \caption[]{{\it Left panel:} Density distribution of \gaia red giant
    branch stars in the Northern galactic hemisphere. The stars were
    selected based on color-magnitude and proper motions to be likely
    Orphan members. {\it Right panel:} Model of the stellar density including foreground and the stream.}
   \label{fig:gaia_track}
\end{figure*}

Figure~\ref{fig:gaia_track} presents the view of the OS using the GDR2
red giants (RG) and BHB stars with $G<18.5$ after applying the
color-magnitude mask shown on Fig.~\ref{fig:cmd_gaia} and the proper
motion selection described above.  The left panel of the Figure shows the
density of the selected candidates in the stream-aligned coordinates
$\phi_1$ and $\phi_2$. We only show the stream in the northern Galactic
hemisphere, as it is harder to identify in the South. The stream is
readily recognizable in the region $20\degr<\phione<120\degr$. Below
$\phi_1\sim20^{\circ}$, the Galactic foreground starts to dominate and
thus the stream is harder to spot despite it being closer to the the
Sun, while above $\phi_1\sim120^{\circ}$, the stream either moves too
far away, so the stellar density drops below detectable levels, or it
stops altogether. Note that the estimates of the Northern extent of
the stream agree well between RR Lyrae and RGBs.

To extract the track of the stream as well as its density and width we
construct a stellar density model that we fit to the data shown on the
left panel Fig.~\ref{fig:gaia_track}. We follow an approach similar to
that adopted in \citet{Erkal2017} to describe the Palomar 5
stream. More specifically, the properties of the stream and the
background are described by cubic splines, that are functions
of the angle along the stream.  The full model for the stellar density
in a given spatial position is as follows:

\begin{equation}
\begin{split}
\rho(\phione, \phitwo) = \exp \left[ {\mathcal B}(\phione) + \phitwo\  {\mathcal B_1}(\phione) + \phitwo^2\ {\mathcal B_2} (\phione)\right]  +\\
 \exp [{\mathcal I}(\phione)] \exp \left( -\frac{1}{2}\left[\frac{\phitwo-\Phi_2(\phione)}{\exp[{\mathcal S}(\phione)]}\right]^2 \right)
\end{split}
\label{eq:spline}
\end{equation}

\noindent where ${\mathcal B}(\phione)$, ${\mathcal B_1}(\phione)$ $ {\mathcal
  B_2} (\phione)$, ${\mathcal I}(\phione)$, ${\mathcal S}(\phione)$, ${\Phi}_2(\phione)$ are
the splines for the logarithm of the background density, the slope of log-background
across the stream, the quadratic term for the log-background, the logarithm of stream's
central stellar density, the logarithm of the stream width and stream track on the sky respectively.
The parameters of the model are the values of the spline
at the spline nodes/knots. In contrast with \citet{Erkal2017} we do not adjust the location of the knots, so they are spaced equidistantly from $\phione=12.5\degr$ to $\phione=121.5\degr$.  To model the \gaia data we adopt a
spline with 7 knots for the density, 5 knots for the stream width, 8 knots for the stream track and  7 knots for the background density and the slopes. The data used in
the modeling are the number counts of stars in $1\degr \times 0.1\degr$
wide bins in $\phione, \phitwo$, while the likelihood function is the
Poisson likelihood for the number counts given the density model.  The
model is implemented in Stan probabilistic programming language
\citep{Carpenter2017}. The only nontrivial priors used are those on the
stream width values at the knots: ${\mathcal N}(\log 0.9, 0.25)$ and stream track: ${\mathcal N}(0,2.5)$. The posterior of the model is sampled by Stan software using Hamiltonian Monte-Carlo No-U-Turn-Sampling algorithm  \citep{MCMCBook,Hoffman2014,Betancourt2017}. We use 15 parallel chains that ran for 2000 iterations each. The convergence of the chains is verified by the Gelman-Rubin $\hat{R}<1.1$ diagnostic \citep{Gelman1992}. When we provide the measurements of individual parameters from the chains, those are medians with 16\%, 84\% error-bars or standard deviations in the case of symmetric posteriors.
We provide the code of the model as
supplementary materials of the paper, as well as on Github\footnote{\url{https://github.com/segasai/orphan_stream_paper}}. The right panel of
Figure~\ref{fig:gaia_track} displays the best-fit model representing
both the stream and the Galactic foreground in the plane of $\phi_1$
and $\phi_2$. The stream model is able to reproduce the observed OS
properties over the entire range of the stream longitudes
considered. In particular, in both the data and the model, the stream
can be seen running at a roughly constant $\phitwo$ for
$20^{\circ}<\phione<50^{\circ}$. The OS then climbs up from negative
$\phitwo$ at $\phione\sim50^{\circ}$ to $\phitwo\sim4^{\circ}$ around
$\phione\sim90^{\circ}$. From there onwards, the stream appears to
broaden and gently slope down for
$90^{\circ}<\phione<120^{\circ}$. Table~\ref{tab:gaia_rgb_fit}
provides the stream track measurements from the model.

\begin{figure*}
\includegraphics[width=1\textwidth]{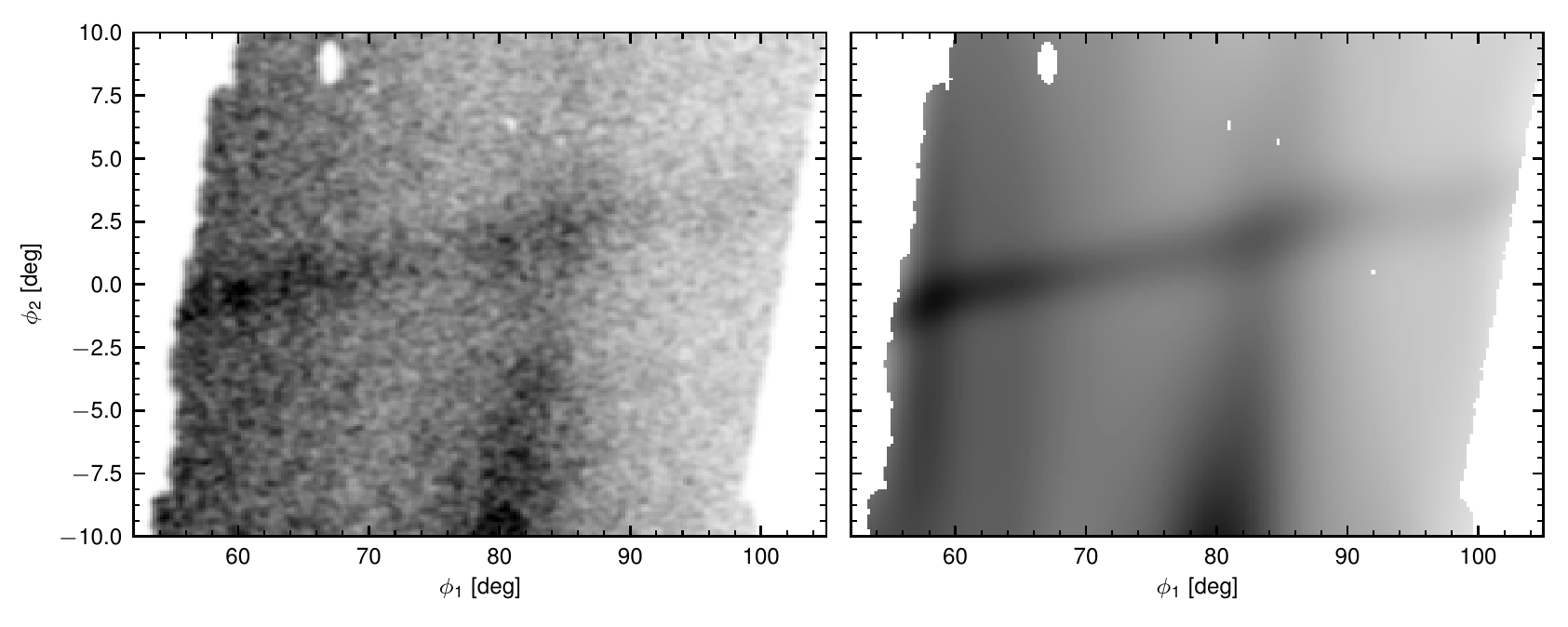}
\caption{{\it Left panel:} The stellar density of matched filter
  selected stars near Orphan stream. We used an old metal-poor
  PARSEC isochrone shifted to the distance of Orphan at each $\phi_1$ to
  define the matched filter. The Orphan stream spans the field nearly horizontally, crossing the almost vertical Sagittarius stream at $\phione \sim 80\degr$.
  {\it Right panel:} The best fit stellar density model of the stream.}
\label{fig:decals_track}
\end{figure*}

The stream can also be revealed and measured in the deeper DECaLS
imaging. The DECaLS DR7 imaging footprint covers the stream from
$50\degr\lesssim \phione \lesssim 100\degr$ at survey depth of $r\sim 23$. Note
however, that to extract the OS signal at the magnitudes fainter than
the \gaia's limit, where the proper motion information is not
available, we have to resort to a matched-filter approach \citep[see
  e.g.][]{Rockosi2002}. In particular, we use the $12$\,Gyr PARSEC
isochrone \citep[see][]{parsec} with [Fe/H]$=-1.5$ and the RR Lyrae
distance track to describe the color-magnitude distribution of the
stream stars. To define the best color-magnitude filter we split the
whole DECaLS field into 100 bins of $\phione$ and construct the best
binary CMD mask for a given $\phione$. We remark that, similarly to
\citet{Erkal2017}, we use an optimal binary CMD mask instead of using
the weights - defined in \citet{Rockosi2002} - to preserve the Poisson
distribution of the filtered map.  Here we use all the sources from
DECaLS DR7 that were classified as PSF (\texttt{type}='PSF') and have $r<23$.
Figure~\ref{fig:decals_track} shows the resulting stellar density
distribution, where the stream can be traced over $\sim45^{\circ}$. In
this particular slice of the OS, the stream seems to be following
closely a great circle without significant curvature.  The vertical
structure at $\phione\sim 80^{\circ}$ is the Sagittarius stream. We
remark that the stellar density along the stream cannot be interpreted
directly as the color-magnitude matched filter mask is different for
different $\phione$.  The right panel of the Figure presents the
spline model of the stellar density shown in the left panel. The
functional form for the model is the same as used for the \gaia RGB
stars (Eq.~\ref{eq:spline}) but has a different number of knots (7 for
the stream width and intensity, 10 for the stream track and 15 knots
for the background and slopes).  The measurements of the stream track
are recorded in Table~\ref{tab:gaia_decals}.

\begin{figure}
  \centering \includegraphics{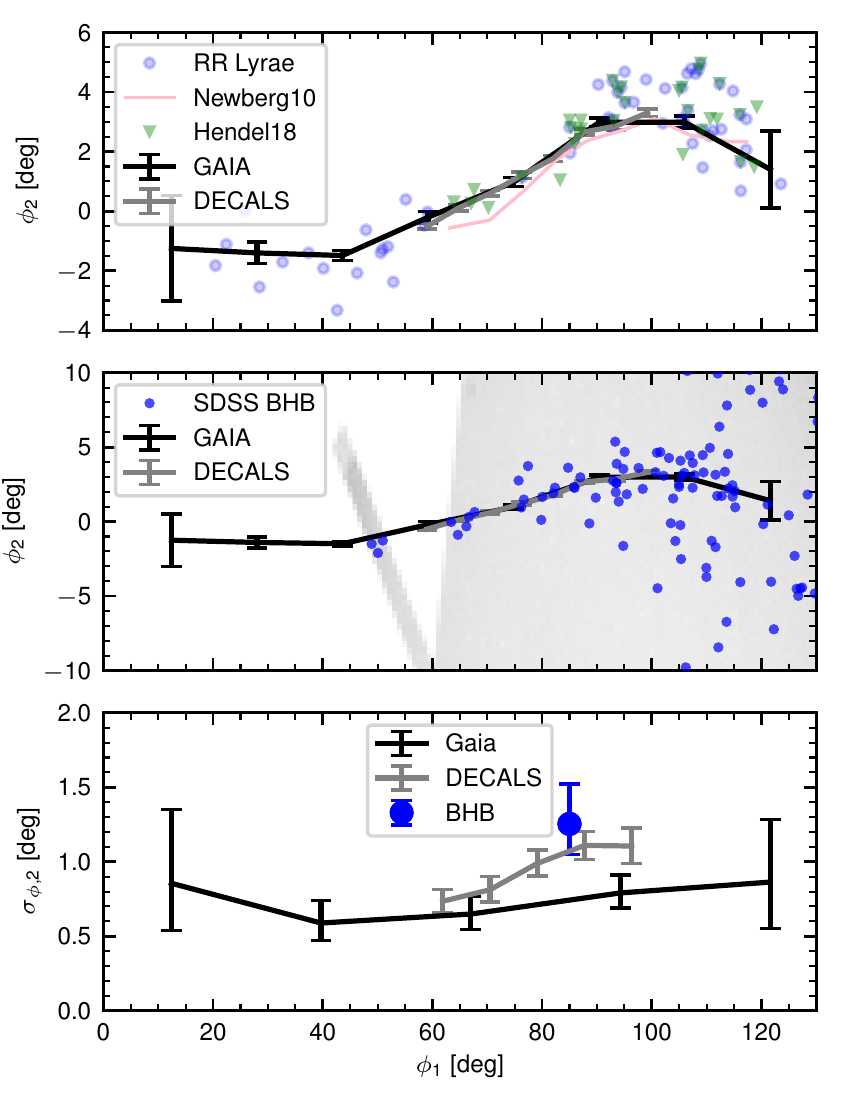}
  \caption[]{Measurements of the Orphan stream in the North using different tracers.
  {\it Top panel:} The track of Orphan stream as traced
    from by RR Lyrae in \Gaia, RGB in \gaia and MSTO and RGB stars in
    DECaLS survey. RR Lyrae are shown by blue points. The black line
    with error-bars shows the extracted track from \gaia RGB stars. The
    grey line shows the stream track from DECaLS data.
    We also show the measurements from \citet{Newberg2009} by a pink line and
    potential Orphan RR Lyrae from \citet{Hendel2018} by green triangles.
    {\it Middle panel:} The Orphan stream as traced by BHB stars. Blue circles show
    BHB stars selected using SDSS $u$, $g$, $r$ photometry, proper motion and distance.
    The grey shading in the background shows the SDSS footprint. We also overplot the stream
    tracks from DECaLS and \Gaia.
    {\it Bottom panel:} The stream width (Gaussian $\sigma$) as a function of
    angle stream longitude, as measured from \gaia RGB stars (black
    line) and DECaLS (grey line) and BHBs (a single blue point).}
   \label{fig:compare_track}
\end{figure}

Figure~\ref{fig:compare_track} summarizes the measured properties of
the stream in different tracers as well as compares them with previous
measurements.  The top panel shows the stream tracks on the sky from
the GDR2 RGs (solid black) and the DECaLS matched-filtered data (solid
grey). Blue circles show individual GDR2 RR Lyrae selected in
Section~\ref{sec:rrl}.  It is reassuring to see that the GDR2 RG and
DECaLS $\phi_2$ centroids of the tidal debris agree perfectly with
each other and with the on-sky positions of the GDR2 RR
Lyrae. Interestingly, for $70^{\circ}<\phi_1<100^{\circ}$, both GDR2
RG and DECaLS tracks show a small offset downwards with respect to the
RR Lyrae locations.  However, this mismatch is modest (smaller than
the stream width).  In the middle panel we show the comparison of the
measured tracks with another tracer, this time Blue Horizontal Branch
(BHB) stars. We select these BHB stars from the SDSS using the
standard $u-g$, $g-r$ color criteria \citep{Sirko2004,Deason2011},
proper motion $|\muone - \hat{\mu}_{\phi,1}(\phi_1)|<0.75$,
$|\mutwo|<0.75$ and distance
$|g-M_G(g-r)-5\log_{10}(\hat{D(\phi_1)})+5|<0.3$ cuts (where
$M_G(g-r)$ is the color-absolute magnitude relation from
\citet{Deason2011}). We note that the BHBs seem to match the other
measured tracks, suggesting that the reason for the mismatch with RR
Lyrae could be either some incompleteness in the catalog or random
sampling effects.  The bottom panel of the
Figure~\ref{fig:compare_track} shows the measured widths from \gaia,
DECaLS and SDSS BHB stars as a function of angle along the
stream. Curiously, there seems to be a slight inconsistency in the
stream width behavior. In the same region of the sky, i.e
$70^{\circ}<\phi_1<100^{\circ}$, the DECaLS-based measurements yield a
slightly wider stream than deduced from the GDR2 RG sample. The BHBs
are also pointing at a higher stream width.  One possible
explanation for the minor inconsistencies in the stream track
measurements with different stellar tracers could be the strong
variation of the foreground in this portion of the sky. As
Figure~\ref{fig:decals_track} demonstrates, the region with
$70^{\circ}<\phi_1<90^{\circ}$ is affected by the presence of the
prominent and broad Sagittarius tidal stream. An alternative
explanation is instead that the Orphan stream has different stellar
sub-populations that also have different stream widths. That scenario
would naturally occur if the disrupting system had components with
different sizes and velocity dispersions.

\begin{table}
\caption{Stream track measurements from \gaia RGB stars}
\begin{center}
\begin{tabular}{ccc}
\hline
$\phi_1$  & $\phi_2$ & $\sigma_{\phi,2}$ \\
deg & deg & deg\\
\hline
12.4 & -1.2511 & 1.7699 \\
28.0 & -1.4023 & 0.3595 \\
43.6 & -1.4896 & 0.1656 \\
59.2 & -0.2057 & 0.1936 \\
74.8 & 0.9761 & 0.1461 \\
90.4 & 2.9812 & 0.1412 \\
106.0 & 2.9882 & 0.2027 \\
121.6 & 1.4035 & 1.2939 \\
\hline
\end{tabular}
\label{tab:gaia_rgb_fit}
\end{center}
\end{table}
\begin{table}
\caption{Stream track measurements from DECaLS matched filtered stars}
\begin{center}
\begin{tabular}{ccc}
\hline
$\phi_1$  & $\phi_2$ & $\sigma_{\phi,2}$ \\
deg & deg & deg\\
\hline
53.275 & -1.6361 & 0.3234 \\
59.0194 & -0.5059 & 0.0881 \\
64.7639 & 0.1056 & 0.0955 \\
70.5083 & 0.5982 & 0.0964 \\
76.2528 & 1.203 & 0.1101 \\
81.9972 & 1.7661 & 0.0973 \\
87.7417 & 2.6617 & 0.1034 \\
93.4861 & 2.8583 & 0.1355 \\
99.2306 & 3.3051 & 0.1291 \\
104.975 & 3.5328 & 0.5728 \\
\hline
\end{tabular}
\label{tab:gaia_decals}
\end{center}
\end{table}

\subsection{The Stream CMD}

\begin{figure}
\includegraphics{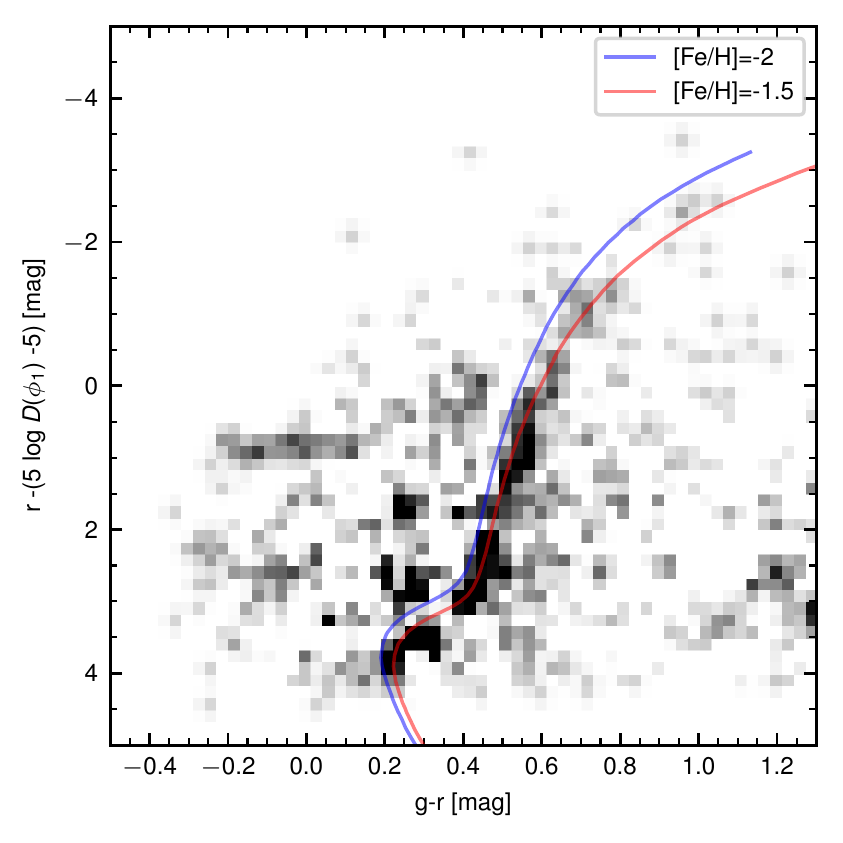}
\caption{The background subtracted absolute magnitude Hess diagram of
  the Northern Orphan stream. We use the combination of DECaLS DR7 and
  \gaia DR2 data. The stars were selected in the area
  $60^{\circ}<\phione<98^{\circ}$, based on \gaia\ proper motion
  $|\mu_{\phi,1}-{\hat \mu}_{\phi,1}(\phione)|<1$, $|\mu_{\phi,2}|<1$ and proximity
  to the stream track determined from DECaLS data $|\phi_2 - \Phi_{2}(\phione)|<1.5\degr$. The background stars were selected in the region $2\degr<|\phi_2 - \Phi_2(\phione)|<5\degr$ The magnitudes
  are corrected by the distance modulus model determined from RR Lyrae
  (see Fig.~\ref{fig:selection}). We over-plot two old 12\,Gyr PARSEC isochrones for
  [Fe/H]=$-2$ and $-1.5$.  }
\label{fig:cmd_decals}
\end{figure}

Having clarified the distribution of the OS stars on the sky, we can
now refine the stream's signal in the CMD using the combination of
\gaia\ and DECaLS data. We use the region with
$60^{\circ}<\phione<98^{\circ}$ and
$|\phitwo-{\Phi_{2}}(\phione)|<1^{\circ}$ to select stream stars
and $60^{\circ}<\phione<98^{\circ}$ and
$2^{\circ}<|\phitwo-{\Phi_2}(\phione)|<5^{\circ}$ region for
the foreground CMD. Here the $\Phi_2(\phi_1)$ is the spline
stream track model determined from DECaLS/\Gaia\ data, rather than \Gaia\ RR
Lyrae (see Table~\ref{tab:gaia_decals}). We apply proper motion
selection cuts base on RR Lyrae proper motion track $|\muone -
\hat{\mu}_{\phi,1}(\phione)|<1$, and $|\mutwo|<1$ and shift each star
by the distance modulus measured from RR Lyrae $5
\log(\hat{D}(\phione))-5$. The stream's Hess diagram (density of stars
in the CMD space) in DECam filter system is shown in
Figure~\ref{fig:cmd_decals}.

While \gaia comfortably detects the stream RR Lyrae and Red Giant stars,
its photometry is too shallow to reach the Main Sequence Turn-Off (MSTO) at
the typical distances of the OS tidal debris. The addition of the
deeper DECaLS data allows us to reach the MSTO and thus break the
distance-metallicity degeneracy. Dark regions (corresponding to
over-dense portions of the Hess diagram) in
Figure~\ref{fig:cmd_decals} reveal several familiar sub-sets of a
co-distant and co-eval stellar population. The most prominent feature
here is the Red Giant Branch, but a tight Horizontal Branch is also
clearly visible, attesting to the quality of the distance
measurements. Less obvious, but also discernible are hints of the
Asymptotic Giant Branch and perhaps even the Blue Stragglers. Given
the presence of the MSTO and with an accurate distance measurement in
hand we can overlay the appropriate isochrone. As evidenced from the
Figure, an old and metal-poor ($[{\rm Fe/H}]=-2$, blue line) isochrone is too
bright for the Stream's MSTO and too steep for its RGB. An isochrone
with $[{\rm Fe/H}]=-1.5$ (red line) does a better job at describing the OS
  stellar populations.

Going back to Figure~\ref{fig:cmd_gaia}, we can compare the CMDs
constructed separately for the Northern (left) and Southern (right)
portions of the Stream. These Hess diagrams use \Gaia ~data only and
apply cuts on the star's proximity to the OS track on the sky and in
the proper motion space (see previous section). Limited to the
\Gaia-only and North-only data, the Stream's CMD appears tidier than
that shown in Figure~\ref{fig:cmd_decals}, albeit reaching only the
bottom of the RGB. The CMD of the Southern part of the stream (right
panel), on the other hand, looks clearly worse, most likely due to
higher levels of Galactic extinction and the more severe foreground
contamination. Nonetheless, the two Hess diagrams seem to share
several features in common: the brighter portions of the RGB, as well
as hints of the AGB and the HB. Given the large amount of noise in the
Southern OS CMD, we can only conclude that the stream stellar
populations below and above the Galactic disc are roughly consistent
with each other, though slight differences are still permitted by the
data in hand.  This motivates a spectroscopic follow-up campaign to
obtain detailed chemistry for RGB stars across the full extent of the
stream to place stronger limits on stellar population variations over
the entire length of the OS.

\subsection{The Stream radial velocities in the SDSS} \label{sec:vrad}

\begin{figure}
\includegraphics[width=0.49\textwidth]{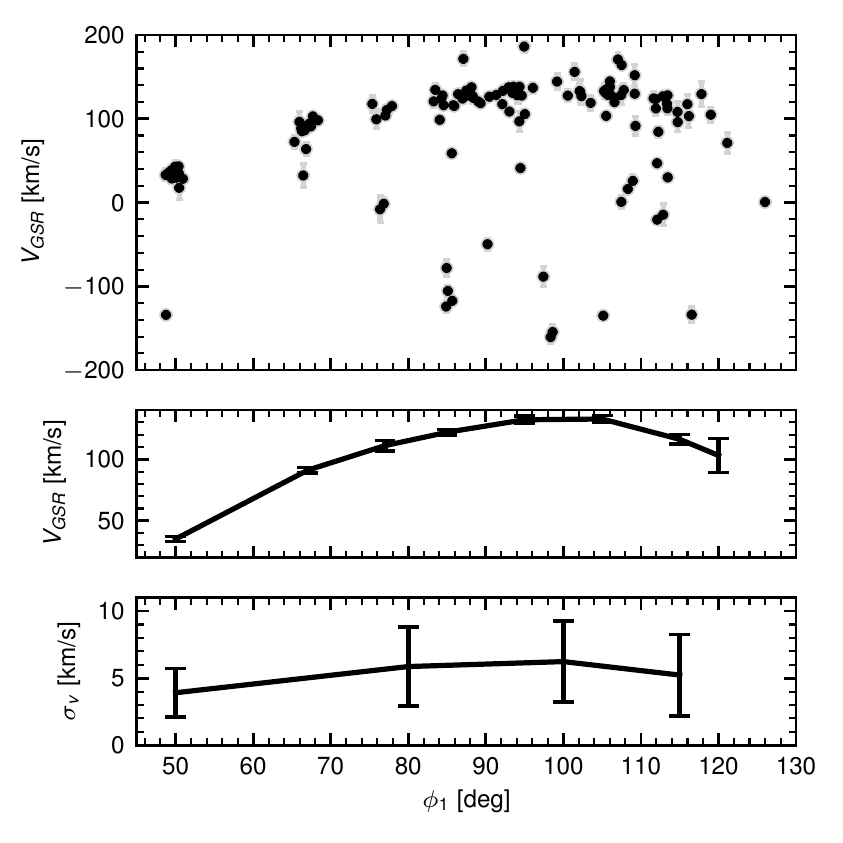}
\caption{
{\it Top panel:} Heliocentric radial velocities of the Northern stream stars
  selected by the position on the sky (within 1.5 degrees of the OS
  track as determined from \Gaia's RGBs), proper motion and their
  colors and magnitudes. Note that the radial velocities are
  approaching zero near $\phi_1 \sim 30\degr$, where the heliocentric
  distances reach the minimum (as seen on Fig.~\ref{fig:selection})
{\it Middle panel:}  Radial velocity measurements extracted from the top panel using
the spline model. The values and error-bars show the mean and standard deviation of
  velocities at the spline knots.
{\it Bottom panel:} The velocity dispersion measurement along the stream.
 The error-bars are 16\% and 84\% uncertainties.}
\label{fig:hrv}
\end{figure}

This sub-section complements the 5-D measurements of the OS track with
a look at the stream's line-of-sight velocities identified in the SDSS
data. We cross-match the SDSS stars with SSPP parameter measurements
\citep[][]{Lee2008,AllendePrieto2008,Yanny2009, Ahn2012} with the
GDR2.  Then we select the likely OS member stars according to their
position on the sky $|\phitwo - {\Phi_2}(\phione)|<1^{\circ}$ using the
\gaia RGB track, proper motion $|\mu_\phi,1-\hat{\mu}_{\phi,1}(\phione)|<1$\,mas\,yr$^{-1}$ and their position in CMD space, using the CMD mask shifted to the distance of the stream $\hat{D}(\phione)$ at each $\phi_1$.  We also require the metallicity to be
$[\textrm{Fe}/\textrm{H}] < -1$ and the star to be classified to be above the Main Sequence
$\log g < 4$. Top panel of Figure~\ref{fig:hrv} shows the line-of-sight
velocities (corrected for the Solar reflex motion) of the likely OS
stars as a function of the angle along the stream. The stream radial
velocity signal is obvious and consistent with previous
measurements \citep[see e.g.][]{Newberg2010}. To take care of the
contamination and noticeable uncertainties in radial velocities, we
model this velocity distribution with a mixture of Gaussians where the
behavior of the mean stream velocity, the stream width in velocity as
well as the mixing fraction of the contamination is described by
splines, i.e.:

\begin{equation}
\begin{split}
{\mathcal P}(V|\phione,\theta) = f_{bg}(\phione)\, {\mathcal N}( V|V_{bg},\sigma_{bg})+\\
(1-f_{bg}(\phione))\, {\mathcal N}( V|V_{str}(\phione),\sigma_{str}(\phione))
\end{split}
\end{equation}
here $\mathcal{N}()$ is the normal distribution and $\theta$ is the
shorthand for all model parameters. $V_{bg}$, $\sigma_{bg}$ are the
mean velocity and the velocity dispersion of the contaminating
population (predominantly the MW stellar
halo). $V_{str}(\phione)$, $\sigma_{str}(\phione)$ are the mean velocity
and the velocity dispersion of the stream. $f_{bg}(\phione)$ is the
fraction of the background stars that is parameterized as a
logit-transform of a spline. The model is implemented in STAN and the
code is provided as supplementary materials. The locations of spline
knots for the velocities together with the measurements of velocities
at the knots are given in Table~\ref{tab:rv_meas}. We show the
measurements in the middle and bottom panels of Figure~\ref{fig:hrv}.
Unsurprisingly, the RV measurements display the same trend as the one
discernible in the top panel of the Figure. The velocity dispersion
measurement, on other hand, shows roughly constant value throughout the
stream of $\sim$ 5 km\,s$^{-1}$. Therefore, it is likely that the
increased apparent scatter in the top panel of Figure~\ref{fig:hrv} at
high $\phi_1$ is mostly due to higher random RV errors of the fainter
stream stars.

\begin{table}
\caption{Measurement of Orphan stream velocity track from  SDSS spectroscopic observations. The third column in the table is radial velocity measurement uncertainty rather than velocity dispersion.}
\begin{center}
\begin{tabular}{ccc}
\hline
$\phi_1$ & $V_{GSR}$ & $\sigma_V$ \\
deg & km\,s$^{-1}$ & km\,s$^{-1}$ \\
\hline
50 & 35.097 & 2.042 \\
67 & 91.084 & 2.258 \\
77 & 111.054 & 4.072 \\
85 & 121.997 & 2.296 \\
95 & 132.123 & 3.092 \\
105 & 132.763 & 2.741 \\
115 & 116.263 & 3.964 \\
120 & 103.205 & 13.798 \\
\hline
\end{tabular}
\end{center}
\label{tab:rv_meas}
\end{table}
\begin{table}
\caption{Adopted light curve periods and Bailey types for the likely Orphan RR Lyrae members. The full
  version of the table is available online.}
\begin{center}
\begin{tabular}{ccll}
\hline
source\_id & Period (d) & Type & Reference \\
\hline
  3890405535908322816 & 0.772158 & RRab & (2)\\
  6417782236698088576 & 0.729023 & RRab & (1)\\
  6572607831962125056 & 0.601812 & RRab & (1)\\
  642802947461337728  & 0.542886 & RRab & (2)\\
  6561357319748782976 & 0.668472 & RRab & (1)\\
  6459293778511267200 & 0.590353 & RRab & (1)\\
  5793432594747653376 & 0.565820 & RRab & (1)\\
\multicolumn{4}{l}{...}\\
\hline
\end{tabular}
\end{center}
\label{tab:rr_periods}
\raggedright \textsc{References:} (1)= \citet{Clementini2018}, (2)=\citet{Sesar2017b}, (3)= This Work\\
\end{table}

\subsection{The population of RR Lyrae stars}

In this section we discuss the properties of the RR Lyrae population
of the stream, in particular in terms of their Oosterhoff types I or
II \citep{Oosterhoff1939}, defined by the location of the stars in two
separate loci in the Period-Amplitude diagram. The fraction of RR
Lyrae of an Oosterhoff type sets apart globular clusters and dwarf galaxies:
globular clusters exhibit the well-known Oosterhoff dichotomy, with
clusters having a vast majority of their RR Lyrae either of Oosterhoff
type I (OoI) or of type II (OoII); while dwarf galaxies commonly
exhibit a mix of the two types as well as intermediate type stars
\citep[see e.g.][]{Catelan2009}.

For the Oosterhoff classification, the light curve period is
required. Out of the \nmem\ kinematically selected RR Lyrae OS
members, only the 86 identified by the Specific Objects Study (SOS) 
\gaia pipeline \citep{Clementini2018} have periods reported in
the \texttt{vari\_rrlyrae} table. Out of the remaining stars, we
retrieved the periods for the 14 stars with matching bona
fide\footnote{PS1 candidate RR Lyrae with classification scores
  $s_3>0.8$ for type ab or $s_3>0.55$ c, as suggested by
  \citet{Sesar2017b}.} counterparts in the PS1 RR Lyrae catalogue from
\citet{Sesar2017b}. Finally, for the remaining stars we retrieved the
$G$, $G_{\mathrm{BP}}$ and $G_{\mathrm{RP}}$ epoch photometry from the
\gaia archive and obtained the light curve periods using the
\citet{Lafler1965} method as adapted in \citet{Mateu2012} to consider
the information in multiple bands simultaneously. Out of the 9 stars
for which periods were computed, we obtained good quality indicators
($\Lambda>3.5$) for 8 stars, the remaining one having too few epochs
($<12$ per band) for a smooth light curve to be obtained with this
method.  The periods adopted for the RR Lyrae members are reported in
Table~\ref{tab:rr_periods}.

To compute the fraction of RR Lyrae of each Oosterhoff type, we use
the Period-Amplitude locus given in Equation~2 of
\citet{Belokurov2018a} to separate the stars into OoI or OoII,
rescaled to convert the V-band amplitude of that equation to \Gaia's
G-band, using $A_G=0.925A_V-0.012$ as in \citet{Clementini2016b}. For
all stars we take \texttt{range\_mag\_g\_fov} as an estimate for the
amplitude\footnote{For the SOS RR Lyrae, we have checked that we get
  similar results using the more robust \texttt{peak\_to\_peak\_g} as
  an amplitude estimator.}. We do the Oosterhoff classification only
for the RR Lyrae stars of type ab (84 out of \nmem), since the
separation into Oosterhoff types is less clear for the type c (and
doublemode type d) stars.

In the Orphan stream as a whole we find 63\% of the RR Lyrae are OoI
and 37\% are OoII. Separating the sample by Galactic latitude, we find
similar fractions of 64\%(36\%) and 62\%(38\%) respectively for
OoI(OoII) in the northern and southern hemispheres respectively. So,
the population of RR Lyrae stars in the northern and southern parts of
the stream are indistinguishable.  Globular clusters display either
low ($\lesssim0.2$) or high ratios ($\gtrsim0.8$) of OoII to OoI
fractions because of the Oosterhoff dichotomy, so the relatively even
ratio observed ($\sim0.6$) indicates the RR Lyrae population resembles
that of a dwarf galaxy more than that of a globular cluster.

\subsection{Orphan stream progenitor properties}
\label{seq:progenitor_properties}
We can use the observed number of RR Lyrae stars in the
stream to estimate the initial luminosity of the progenitor. For this, we need to first estimate how incomplete our sample might be. Comparing with the RR Lyrae from \citet{Hendel2018}, we find that out of their 32 stars only 24 (75\%) are present in the \Gaia\ RR Lyrae catalogue used here (prior to the kinematical selection of OS members), out which 15 are in our kinematically selected sample. The \citet{Hendel2018} sample comprises high probability RR Lyrae candidate\footnote{and one medium probability candidate} members of the OS identified by \citet{Sesar2013}, and, although it might also be affected by incompleteness itself, it can serve as a reference to estimate the minimum incompleteness affecting the \Gaia\ sample. Thus, we estimate an expected number of 109 RR Lyrae of type ab, based on the 84 observed ones and assuming a (maximum) completeness of 75\%, in agreement with \citet{Hendel2018}’s prediction of $\sim100$ RR Lyrae.

\begin{figure}
\includegraphics[width=0.49\textwidth]{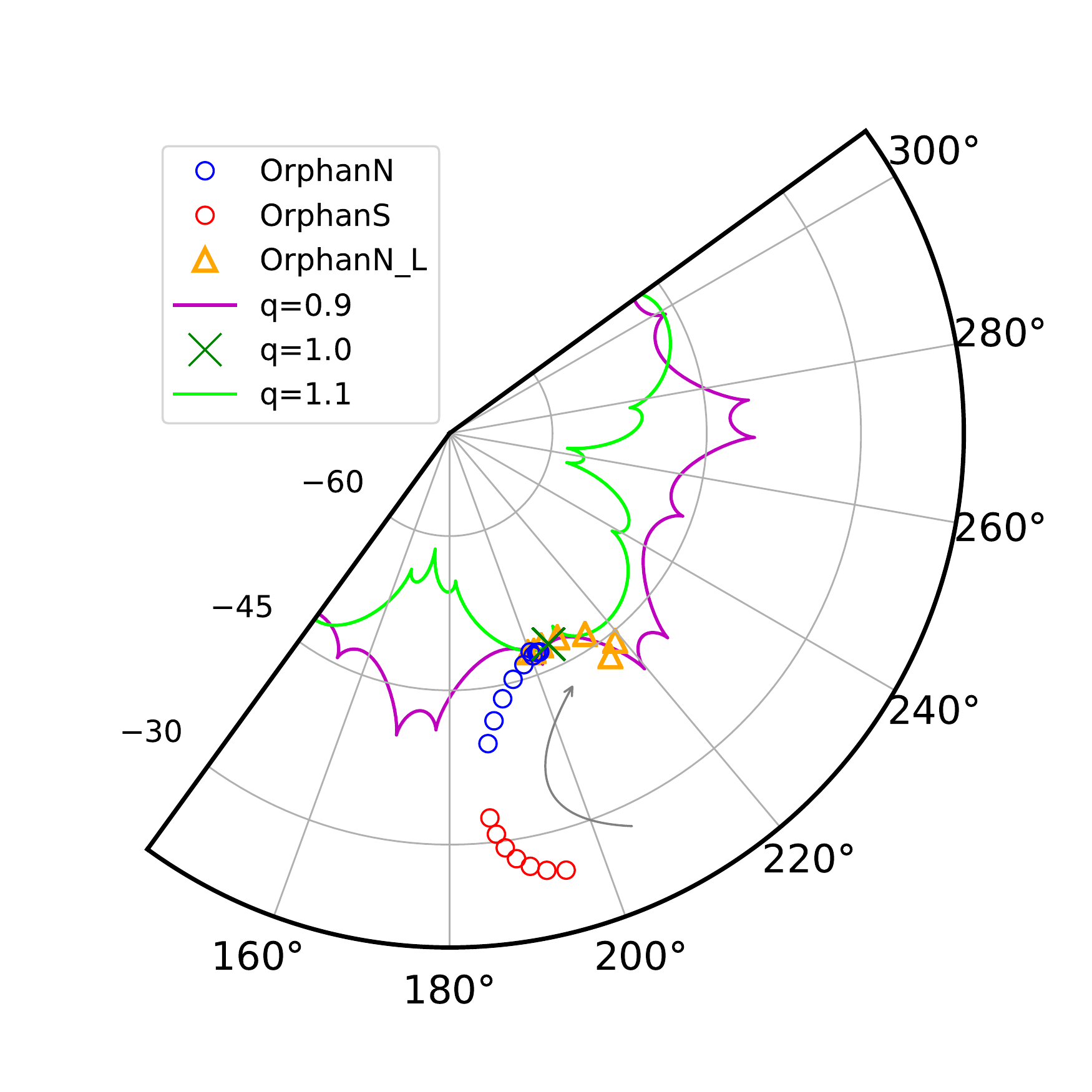}
\caption{Galactocentric orbital pole calculated from the 3-D track of
  individual portions of the Orphan Stream, using the positions of
  likely RR Lyrae stars. The Northern sections are shown in blue
  circles and Southern ones are in red. For part of the Northern
  sections where line-of-sight velocities are available from SDSS, the
  angular momentum pole are also calculated based on full 6D phase
  space information of RR Lyrae stars, and shown as the orange
  triangles. Also  green and violet lines as well as the green cross show  the expected orbital plane precession in
  a logarithmic halo potential with different flattening. The gray
  arrow shows the moving direction of the stream from South to
  North. }
\label{fig:pole}
\end{figure}

We use the inference model as in \citet[][see their
  Sec.~4.3]{Mateu2018a} based on the linear relation observed between
the number of type ab RR Lyrae and the total absolute magnitude $M_V$,
for dwarf galaxies and globular clusters in the Milky Way. For the 109
expected RR Lyrae of type ab, this gives $M_V=-10.8 \pm 1.3$,
corresponding to the posterior mode and 68\% confidence interval,
which translates into an inferred total luminosity with a most
probable value at $L_V=3.8\times 10^5 L_{\sun}$ in the 68\% confidence
interval $0.2$--$3\times 10^6\,L_{\sun}$. The lower and upper bounds
on $M_V$ confidently place the progenitor among the classical dwarfs,
with a luminosity between those of the Sextans and Leo~I dwarf
spheroidals \citep{McConnachie2012}, with the most probable $M_V$
placing it in the top 5 most luminous dwarfs, just close to
Leo~I. This is a robust result, that holds even if no correction for
incompleteness is made, in which case $M_V= -10.4 \pm 1.3$,
i.e. brighter than Sculptor and still in the top 5. Finally, note this
is in agreement with our findings from the Oosterhoff types ratio, that also
point toward a dwarf galaxy progenitor rather than a globular cluster.

As a consistency check, we also use the mass-metallicity relation for Local Group galaxies from \citet{Kirby2013} to estimate a lower-mass limit for the OS progenitor (assuming the OS has a lower metallicity than its host galaxy). Using $[\textrm{Fe}/\textrm{H}] = -1.5$ for the OS, we estimate that the progenitor had a stellar mass of  approximately $4 \times 10^6 \rm{M}_\odot$. This puts its progenitor at roughly between Leo~I and Sculptor, consistent with the luminosity calculation from the number of RR Lyrae above \citep[see also discussion in][]{Hendel2018}. {  We can also compare the velocity dispersion observed in the stream of $\sim$ 5 km\,s$^{-1}$ (see Section~\ref{sec:vrad}) with the stellar velocity dispersion in dwarf galaxies, and that points towards somewhat less massive classical dwarf progenitors like Carina or Leo~II with the velocity dispersions of $\sim 6$ km\,s$^{-1}$ \citep{Walker2007}. The luminosities of these dwarfs however are still consistent with estimates above. }

\begin{figure*}
  \includegraphics[width=0.49\textwidth]{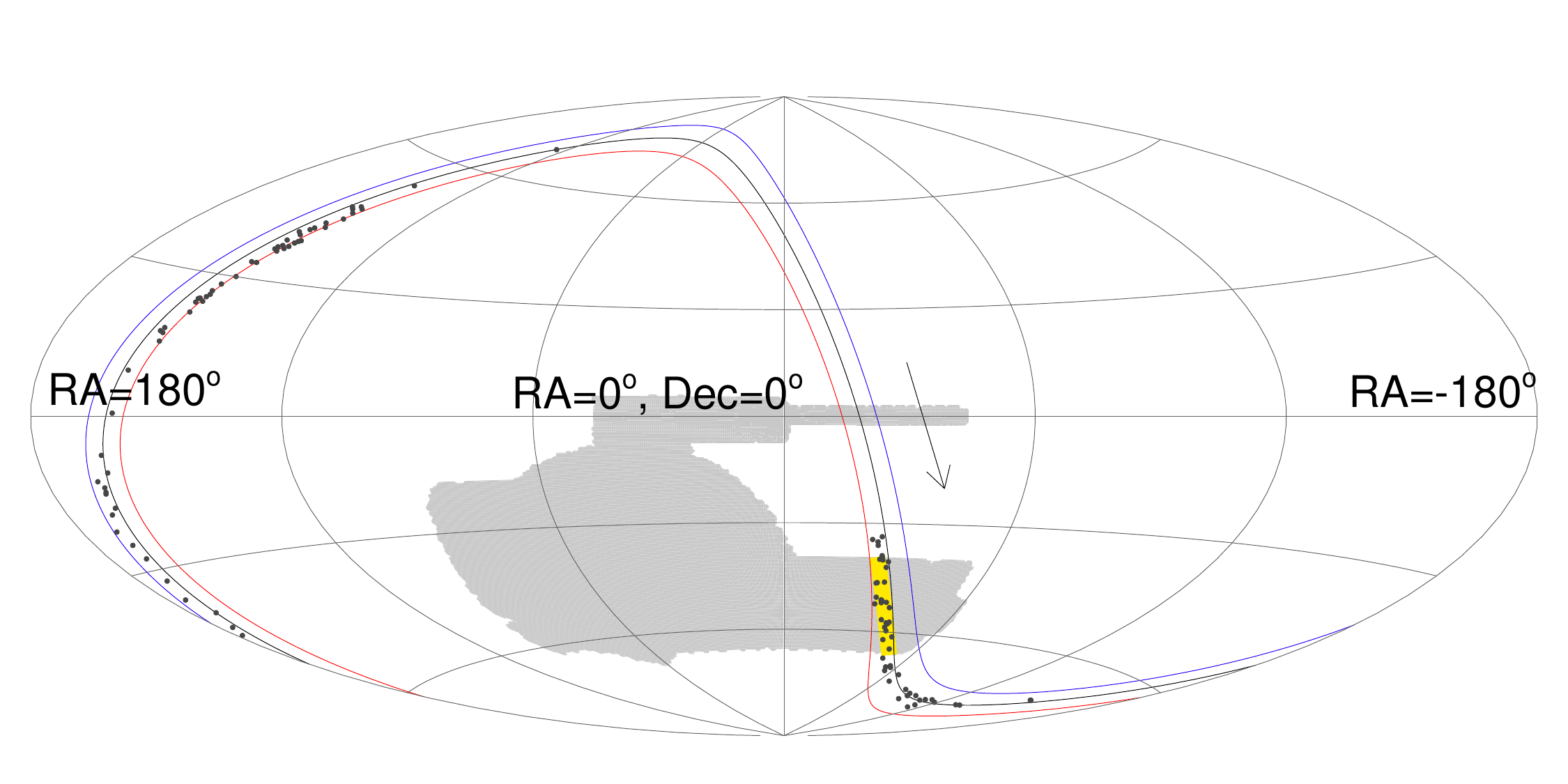}
  \includegraphics[width=0.49\textwidth]{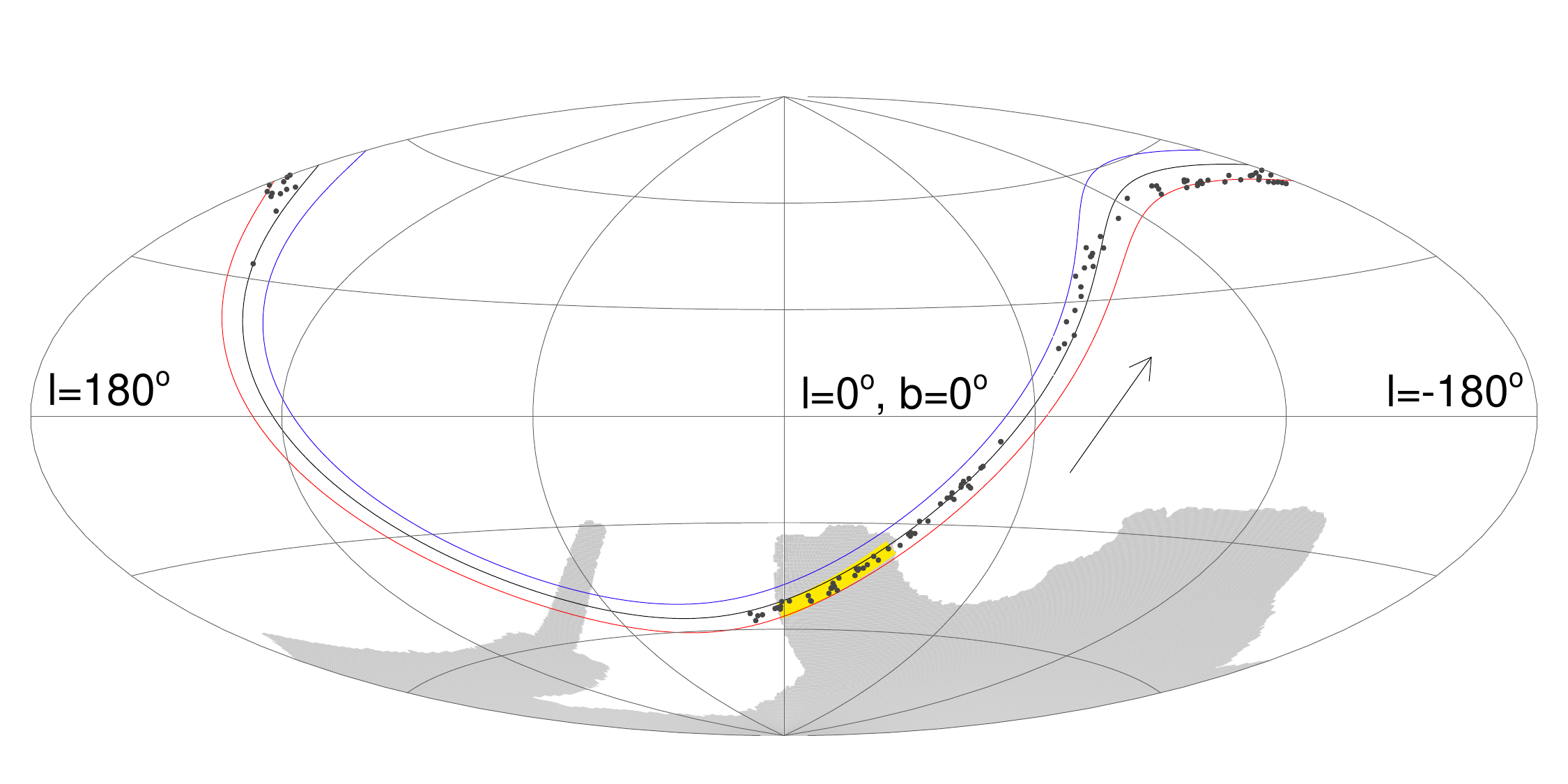}
\caption{Orphan Stream across the sky in Equatorial (left) and
  Galactic (right) coordinates. Grey shaded area marks the boundaries
  of the DES DR1 footprint. Great circles parallel to the equator of
  the OS coordinate system and corresponding to $\phi_2=-4^{\circ}$
  (blue), $\phi_2=0^{\circ}$ (black) and $\phi_2=4^{\circ}$ (red) are
  shown. Arrows indicate the direction of the stream's motion. The
  Chenab track as measured by \citet{Shipp2018} is shown in yellow.}
\label{fig:des_foot}
\end{figure*}

\section{Discussion}
\label{sec:disc}

We have provided an unprecedented view of the Orphan Stream based on
the \gaia DR2 as well as a panoply of other surveys. We have taken
advantage of the GDR2 RR Lyrae; these old and metal-poor pulsating
stars are relatively numerous in the OS, with an average density of
$\sim0.5$ per deg$^2$ and can be traced over $210^{\circ}$ on the sky
reaching distances of $\sim$ 60\,kpc from the Sun on either side of the
Galactic plane. Viewed with the GDR2 RR Lyrae, the OS appears long and
eccentric, piercing the Milky Way's disc at around 10 kpc and
disappearing far into the Galactic halo, more or less in agreement
with earlier studies \citep[see
  e.g.][]{Newberg2010,Sesar2013,Hendel2018}. Unexpectedly, however,
the OS's behaviour changes drastically in the Southern hemisphere: the
Orphan's track on the sky bends, while the debris kinematics appear to
exhibit strong across-stream motion.

\subsection{Orbital plane of the OS debris}

The striking twist of the OS can be comprehensibly demonstrated by
tracking the direction of the vector normal to the orbital plane
passing through individual portions of the stream. First, the
celestial coordinates and line-of-sight distances of the likely OS RR
Lyrae are converted into Galactocentric Cartesian coordinates,
$X,Y,Z$. We then fit each of the three coordinates as function of the
stream longitude $\phi_1$ with a high order polynomial. This smooth
3-D track of the stream is split into 19 reference nodes, separated by
$10^{\circ}$, of which there are 11 in the North and 8 in the
South. The direction of the normal corresponding to the orientation of
the orbital plane passing through the debris between each neighboring
pair of the stream nodes is then simply the cross product of their
position vectors $X_i, Y_i, Z_i$ and $X_{i+1}, Y_{i+1}, Z_{i+1}$.

Figure~\ref{fig:pole} presents the evolution of the OS debris plane in
Galactocentric polar coordinates $l_{\rm GC}, b_{\rm GC}$, where the Northern
sections of the stream are shown in blue and the Southern ones in
red. As the Figure demonstrates, the debris pole swings wildly, by
some $\sim20^{\circ}$, as the stream moves from the South to the
North.
Curiously, the bulk of this orbital plane wobble is in the direction
of the Galactic $b_{\rm GC}$. Such an evolution of the stream's pole is
inconsistent with the orbital plane precession in the potential with a
flattening along the $z$ axis. As highlighted in e.g.
\citet[]{stray}, streams in axi-symmetric potentials are expected to
precess around the symmetry axis. This is illustrated in
Figure~\ref{fig:pole} with two examples of the orbital pole variation,
one for an orbit in an oblate (purple curve) and one in a prolate
(green curve) Dark Matter halo. While the orbital planes do exhibit
small-amplitude nutation along $b_{\rm GC}$, most of the pole
displacement is in the perpendicular direction, i.e. that parallel to
$l_{\rm GC}$ and around the $z$ axis.  Compared to these theoretical
expectations, the OS debris pole exhibits a swing with an amplitude
some 4 times higher. We further compute the direction of the stream's
angular momentum for the Northern stream, where the line-of-sight
velocities are available from the SDSS data (see previous
Section). The direction of the stream's angular momentum and the
debris pole match pretty well in the North. We expect the angular
momentum pole and debris pole to decouple further down South;
furthermore, the line-of-sight velocities of the Southern portions of
the stream (when measured from future surveys) should manifest strong
perturbations indicated by the wobble of the stream track.

\subsection{Orphan Stream in the South}

At first glance, half of our stream detections appear new, as they are
located under the Galactic plane. As illustrated in
Figure~\ref{fig:des_foot}, the Southern portion of the stream crosses
the footprint of the Dark Energy Survey. On closer examination, it is
clear that the Southern part of the Orphan Stream is entirely
consistent with the DES stream Chenab \citep[see][]{Shipp2018}. It is
likely that Chenab stars were not identified as Orphan's due to the
dramatic twist in the OS track on the sky around
$-20^{\circ}<\phi_1<50^{\circ}$ and drift of the orbital pole by more than 20 degrees shown on Figure~\ref{fig:pole}. We leave the detailed analysis of
Chenab-Orphan in the South for a stand-alone publication.

\begin{figure}
\includegraphics{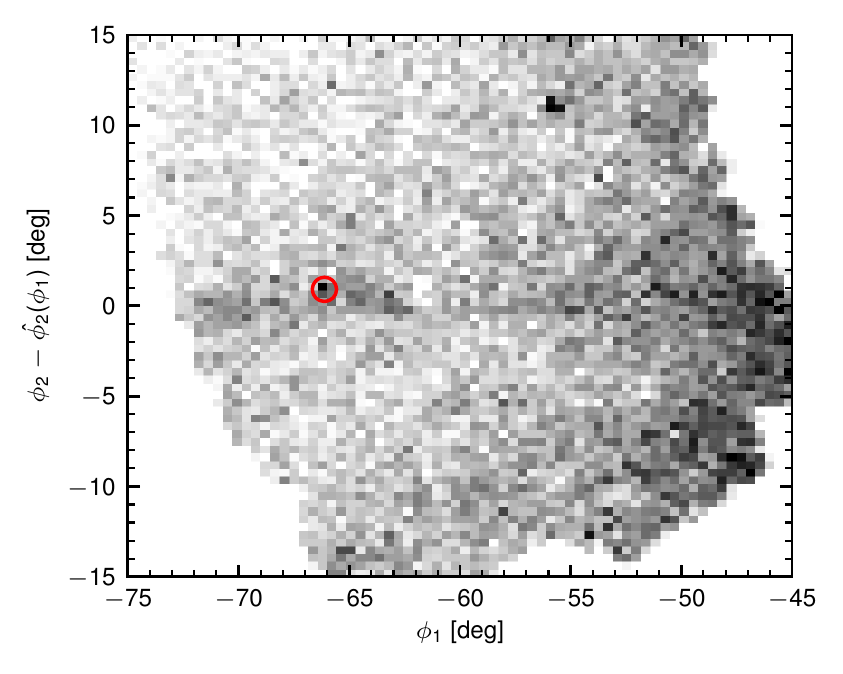}
\caption{Density map of the Orphan Stream in the South (where it was
  previously called Chenab). The figure shows the density of matched filter
  selected MSTO stars from the DES DR1 data. We used the same
  $\phione$-dependent matched filter based on distance, as for
  Figure~\ref{fig:decals_track}. The y-axis of the plot is the 
  offset in the across the stream coordinate $\phitwo$ with respect to the
  stream track in RR Lyrae. The red circle shows the
  over-density of stars associated with the Grus 2 MW satellite.}
\label{fig:des_track}
\end{figure}

In the South, while the stream could not be easily seen with \Gaia's
RGBs, it is clearly visible in the DES DR1 data. Here we proceed in
the same fashion as with the DECaLS data. More precisely, we construct
a set of matched filters based on the $[{\rm Fe/H}]=-1.5$, 12\,Gyr PARSEC
isochrone shifted to the distance $\hat{D}(\phione)$ predicted from
the RR Lyrae. Figure~\ref{fig:des_track} shows the density of 
matched filter selected stars on the sky, with the y axis being the
residual offset with respect to the RR Lyrae track $\phitwo -
\hat{\Phi_2(\phione)}$, so that the stream is expected to go
horizontally near zero. Traced with the DES data, the stream is
unmistakably present, but shows more complicated morphology (compared
to its Northern section). For example, the stream's signal is
strongest at $\phione<-62^{\circ}$, and then drops noticeably at
$\phione>-57^{\circ}$. However, in the middle of this $\phione$ interval, the
debris density is greatly reduced. It is difficult to say at the
moment whether this is an artifact of the data or an intrinsic
variation of the stream's properties. We also note a prominent compact
stellar overdensity on top the stream at $\phione\sim -66^{\circ}$. This
happens to be the recently discovered MW satellite Gru~2
\citep{DrlicaWagner2015}.

\subsection{Gru 2 and other Galactic satellites}

\begin{figure}
\includegraphics[width=0.49\textwidth]{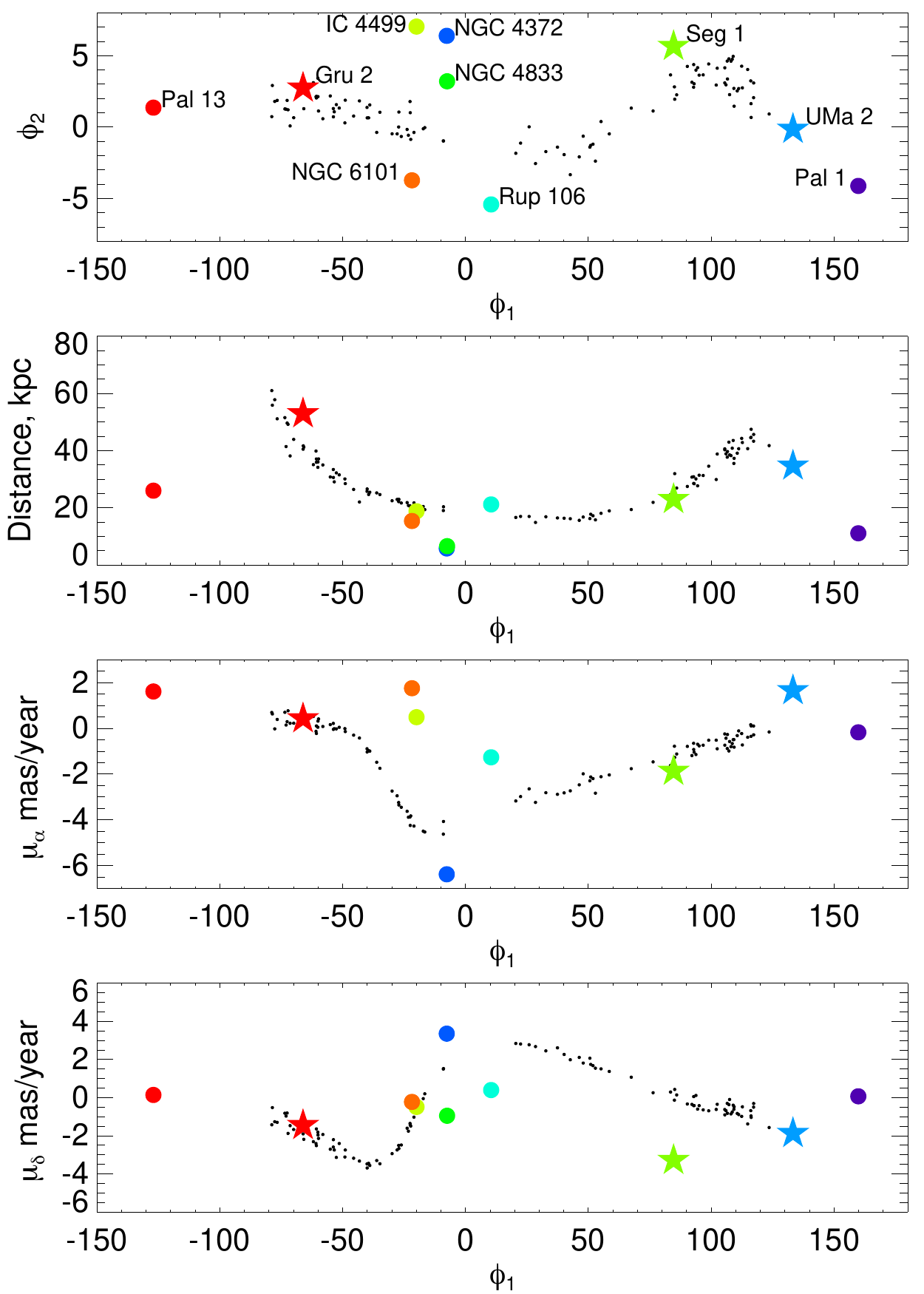}
\caption{Phase-space coordinates of Galactic globular clusters (filled
  circles) and ultra-faint dwarfs with $|\phi2|<7^{\circ}$. GC data is
  from the catalogue of \citet{Vasiliev2018}, UFD data is a
  combination of \citet{Simon2018}, \citet{Fritz2018} and
  \citet{Pace2018}. Small black dots show the positions of the OS RR
  Lyrae.}
\label{fig:sats}
\end{figure}

Figure~\ref{fig:sats} shows the position of Gru~2 (filled red star) in
various phase-space projections, namely in stream-aligned coordinates
$(\phione, \phitwo)$ (top panel), distance $(\phione, D)$ (upper
middle), proper motion $(\phione,\mu_{\alpha})$ (lower middle) and
$(\phione, \mu_{\delta})$ (bottom). Gru~2 is coincident with the OS RR
Lyrae in 3 out of 4 dimensions considered: position on the sky and the
two components of the proper motion. It appears offset to slightly
larger distances by some $\sim10$ kpc compared to the stream's debris
at the same $\phione$. Overall, the connection between Gru~2 and the
OS seems rather likely; however, line-of-sight velocity and chemical
abundances are required to verify the reality and nature of that
association.

Also shown in Figure~\ref{fig:sats} are the positions of the Galactic
globular clusters (GCs; filled circles) and ultra-faint dwarfs (UFDs;
filled stars) nearest to the stream. Only 7 GCs are located within
$|\phitwo|<7^{\circ}$ of the stream's great circle. None of the
globulars match well the stream's phase-space track. Of the 7 objects
displayed, Ruprecht 106 is closest across all dimensions, but
nonetheless it still is considerably offset in proper motion,
i.e.  some $\sim 2$ mas\,yr$^{-1}$ in both $\mu_{\alpha}$ and
$\mu_{\delta}$. We also note that given the estimated luminosity of the
Orphan progenitor galaxy (see Section~\ref{seq:progenitor_properties}), 
we would not expect it to host  globular clusters.
 Apart from Gru~2, two additional UFDs can be seen in
the vicinity of the stream, i.e. Segue~1 and UMa~2, both previously
suspected to have some sort of association with OS
\citep[][]{Fellhauer2007,Newberg2010}. Interestingly, UMa 2 sits right
on the continuation of the stream's track after a notable twist at
$\phione\sim100^{\circ}$. Unfortunately, as Figure~\ref{fig:sats}
demonstrates, the satellite is at least 10 kpc closer along the
line-of-sight and has a faster proper motion along RA. UMa~2's radial
velocity is also discrepant: according to \citet{Simon2007} its
V$_{\rm GSR}\sim -33$ km s$^{-1}$, while as evident from
Figure~\ref{fig:hrv}, at $\phione\sim130^{\circ}$, the OS's radial
velocity is unlikely to be lower than 50 km s$^{-1}$. Segue~1 is close
to the stream in distance and in $\mu_{\alpha}$ but is off by
$\sim3^{\circ}$ on the sky. More importantly, its proper motion along
declination is markedly different from that of the OS's tidal debris
at the same $\phione$, as first pointed out by \citet{Fritz2018}. Note
that there is only one classical dwarf spheroidal galaxy (Leo~I) that is projected
close to the stream on the sky. However, given that Leo~I is located
much further away, at $\sim250$ kpc, it is not shown in
Figure~\ref{fig:sats}.


\subsection{A gap in the Northern section of the Stream}

Stellar streams have been recently been put forward as excellent
detectors of Dark Matter subhalos \citep[see
  e.g.][]{Ibata2002,Johnston2002, Siegal2008, Carlberg2009,
  Yoon2011}. If a subhalo flies near the stream, it can impart a small
velocity kick to the closest stars. The affected stars change their
orbits slightly, and after some time, a region near the point of the
subhalo's closest approach will be evacuated, thus producing an
observable stream gap \citep[see][]{Erkal2015}. The size of such gap
is linked to the properties of the perturber and the fly-by, and is
directly proportional to the mass of the subhalo and the time since
the interaction. Crudely, the more massive the deflector is the larger
the gap. However, all gaps need an appreciable amount of time to grow
in depth and hence there is a typical size-scale for a detectable
density depletion \citep[see][]{Erkal2016}. As these authors
demonstrate, when interacting with streams in the MW, CDM subhalos
with masses between $10^5$ and $10^9$ M$_{\odot}$ typically produce
gaps with sizes larger than $1^{\circ}$, with prevalence of gaps with
$5^{\circ}-10^{\circ}$ in size. 

Although \cite{Erkal2016}
focused on globular cluster streams, these subhalo perturbations will
also create gaps in dwarf galaxy streams like the OS. However, due to the 
larger velocity dispersion of the OS ($\sim 5$ km\,$^{-1}$, see Sec. \ref{sec:vrad})
the smaller gaps will be washed out. \cite{Erkal2016} demonstrated that significant
gaps can be created by perturbations with velocity kicks down to one tenth
the velocity dispersion of the stream, corresponding to $\sim 0.5$ km\,$^{-1}$ for the OS.
Assuming a typical flyby speed of $\sim 300$ km\,s$^{-1}$, this translates to a minimum
subhalo mass of $\sim 3\times 10^7 M_\odot$. Using the results of \cite{Erkal2016},
such a subhalo can create deep gaps (factor of 2 depletion) with sizes as small as several degrees.

\begin{figure}
\includegraphics{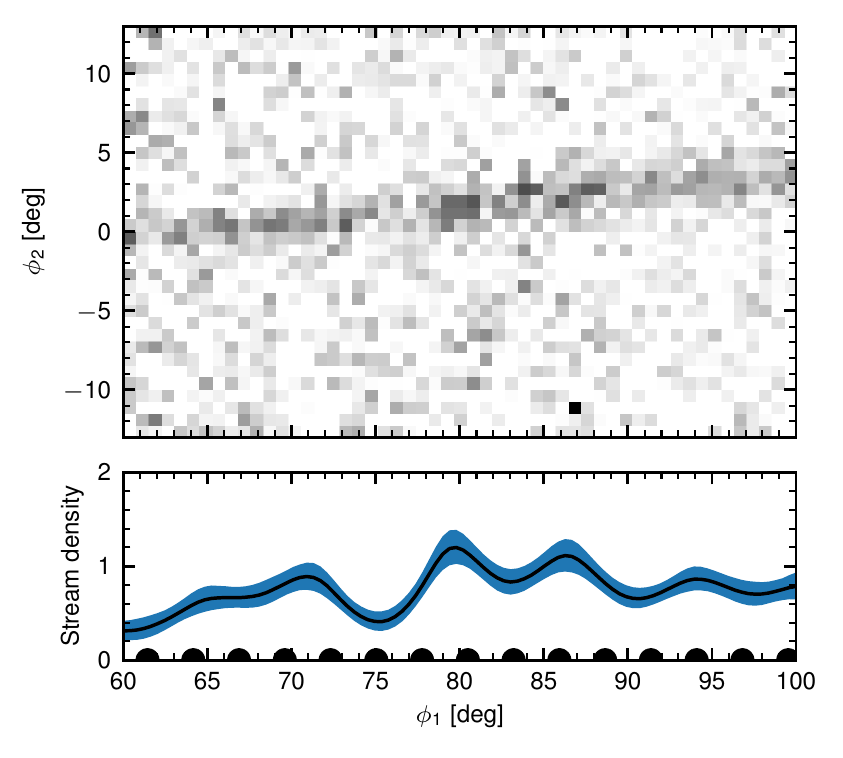}
\caption{
  {\it Top panel: } The density of metal-poor MSTO stars ($0.2<g-r<0.3$ and $3<M_r<4$ in
  the DECaLS data, after subtracting the best-fit background model. A quite prominent
  underdensity is visible at $\phione \sim 76\degr$. We also notice that the stream
  is getting wider at higher $\phione$ confiming measurements
  shown on Figure~\ref{fig:compare_track}. {\it Bottom panel:} The results of fitting the spline model (Eq.~\ref{eq:spline}) to the
   non-background subtracted MSTO density distribution. 
   The curve and the bands are showing the median and 16\%, 84\% percentiles of
   the credible intervals respectively of the linear stream density from the spline model. The black circles identify the location of spline knots for the stream density.
  We see that the model correctly identifies the underdensity at $\phione \sim 76\degr$. The stream density there drops by more than factor of $\sim 2.5$ and	 is highly significant at $>$5\,$\sigma$.}
\label{fig:gap}
\end{figure}

As is clear already from e.g. Figure~\ref{fig:memother}, the number of
OS RR Lyrae changes substantially as a function of stream longitude.
While some of those changes could be caused by real density variations
along the stream, a major factor affecting the density is potential
incompleteness related to the \gaia scanning law.  In fact, the region
of the stream with $50\degr\lesssim \phione\lesssim 90\degr$ is
exactly the area with the smallest number of \gaia observations, with a total number of individual light-curve points around
20, which is near the absolute minimum for the RR Lyrae
identification. Therefore, we think that the paucity of RR Lyrae in
this range could be explained by the Gaia catalog
incompleteness. However, our measurements of turn-off stars with
DECaLS data should not suffer from these issues and should allow us to
detect possible stream gaps.  In fact, looking at
Figure~\ref{fig:decals_track} we see a hint of stream underdensity at
$\phione\sim 75\degr$. However, the interpretation of the Figure is
somewhat complicated by the use of a matched filter that varies
depending on the location along the stream. Figure~\ref{fig:gap}
provides a clearer illustration of the debris density evolution in the
DECaLS data. The top panel shows the density of metal-poor
turn-off stars, $0.2<g-r<0.3$, within a fixed range of absolute
magnitudes, $3<M_r<4$. This image is background subtracted after
fitting the MSTO density map by the spline model (see
Eq.~\ref{eq:spline} and Section~\ref{sec:rgb} for more details). The
2-D density plot shows no clear residuals of the Sagittarius stream,
confirming good background subtraction and reveals a mostly
well-behaved stream with the exception of the area near $\phione\sim
75\degr$ where the stream density seems to noticeably drop. To assess
the significance of this feature we look at the 1-D stream density
profile extracted from the spline model of the MSTO distribution and
shown on the bottom panel of Figure~\ref{fig:gap}. The black line and
the blue band show the median and 16\%, 84\% percentiles of the stream's
linear-density from the posteriors of the spline model (identical to
the one used in Section~\ref{sec:rgb}). The 1-D density profile
confirms the reality of the density gap and allows us to assess its
depth and significance. The stream density in the gap drops by a
factor of 2.5.  According to the uncertainties on the stream density
the gap is highly statistically significant ($>5.5\sigma$) and
has a size (FWHM) of $\sim 4$ degrees. While it is too early to
speculate about the cause of this stream gap, it is consistent with
the gap created by a dark matter halo stream perturbation. We remark
that given the Orphan stream pericenter is quite large $\gtrsim
15$\,kpc, the stream perturbation is less likely to be caused by Giant
Molecular Clouds \citep{Amorisco2016} or the bar
\citep{Hattori2016,PriceWhelan2016}.
\section{Conclusions}
\label{sec:conc}

{ In this paper we present a comprehensive all-sky view of the Orphan stream 
by measuring its track on the sky and in 3-D, constraining stream stars proper motions and radial velocities. This reveals a number of unexpected aspects of the stream's behaviour.}

\begin{itemize}
  \item { We show that the Orphan stream is more than twice as long than previously measured, extending from the Northern galactic hemisphere to the South with the total length of $\sim 210^{\circ}$. The stream is at its closest to the Sun at the distance of $\sim$ 15\, kpc when it crosses the Galactic plane and extends out to distances of $\sim$ 60\,kpc at the stream edges in the South and North.}
  
  \item As a heliocentric observer follows the stream's stars across
    the celestial sphere, the stream's track on the sky exhibits a
    clear systematic shift with an amplitude of several degrees over
    $210^{\circ}$ of the detected OS's length. If the distance to the
    debris is taken into account, the Galactocentric observer would
    see an even more dramatic swinging of the stream's orbital plane,
    i.e. of order $\sim20^{\circ}$. This drastic debris pole
    evolution is mostly in the direction of Galactocentric $b_{\rm
      GC}$ and thus is inconsistent with precession around the
    $z$-axis.

  \item Accompanying the swaying of the debris pole is the strong
    across-stream motion. We show that over several tens of degrees on
    the sky, the stream's stars appear to be moving in the direction
    away from the Orphan's track. Such misalignment between the debris
    proper motion and the stream's direction can not arise if the
    stars orbit in a steady gravitational potential. We therefore
    hypothesize that the evolution of the debris plane and the
    non-zero across-stream motion are connected and are caused by an
    interaction with a massive perturber.

  \item Apart from the large scale perturbation described above and
    limited to $-100^{\circ}<\phione<50^{\circ}$, a smaller amplitude
    twist in the stream track is visible around
    $\phione\sim100^{\circ}$, i.e. at the Northern tip of the
    Stream. Stars in this section do not show a significant
    stream-motion misalignment and thus a different phenomenon may be
    needed to explain this observation.

  \item This paper demonstrates that a portion of the Southern
    extension of the OS had been seen previously. More precisely, a
    stream called Chenab was discovered in the DES data by
    \citet{Shipp2018} which now can be shown to match perfectly the
    GDR2 RR Lyrae detections discussed here. Chenab was likely not connected to the (previously known portion of the)
    OS because of the strong debris plane evolution detected in this paper.

  \item { The analysis of the total population of RR Lyrae shows that the progenitor of the stream was likely a classical dwarf galaxy with the total luminosity of $M_V\sim -10$. }

  \item After many years of searching, a plausible connection between
    OS and a Galactic satellite may have been established. An
    ultra-faint dwarf --- Gru~2 --- overlaps with the Orphan debris in most of
    the phase-space projections, with one small exception: along the
    line-of-sight, the satellite is $\sim10$ kpc further away.

   \item We identify a well-defined density gap in the Orphan stream
     with the size of 4 degrees. The density of the stream in the gap
     drops by a factor of 2.5.  This is consistent with the gap
     created by dark matter substructures with masses in the range of
     $\sim 10^7$M$_\odot$-$10^8$M$_\odot$.

\end{itemize}

This is the first paper in a series by the Orphan Aspen Treasury
(OATs) Collaboration. Note that, released simultaneously with this
work, a companion publication \citep{Erkal2019} presents the results
of an in-depth stream modeling exercise and provides an explanation of
the observed OS properties.

\begin{table}
\caption{The subset of the likely Orphan RR Lyrae selected based on
  their distance, proper motion and position on the sky. The full
  version of the table is available.}
\begin{center}
\begin{tabular}{cccc}
\hline
source\_id & $\alpha$ & $\delta$ & $D_{hel}$ \\
& deg & deg & kpc \\
\hline
3890405535908322816 & 155.36706 & 17.52019 & 32.0 \\
6417782236698088576 & 276.18759 & -74.20531 & 22.5 \\
6572607831962125056 & 329.77628 & -40.60788 & 38.2 \\
642802947461337728 & 150.64721 & 25.24754 & 30.1 \\
6561357319748782976 & 331.32574 & -46.60868 & 41.7 \\
6459293778511267200 & 321.18679 & -58.68641 & 32.2 \\
5793432594747653376 & 225.44879 & -74.39093 & 19.4 \\
6450297166352626048 & 312.65244 & -64.49243 & 26.4 \\
6564567275226088192 & 328.87316 & -46.27519 & 41.2 \\
6459507255565439744 & 321.67634 & -57.45849 & 33.4 \\
3763707471010331520 & 163.83093 & -7.17632 & 20.6 \\
799463292628940672 & 146.00854 & 36.26583 & 40.3 \\
5792708879873126400 & 227.70794 & -74.57514 & 20.5 \\
6425035538810325248 & 311.72339 & -67.05691 & 22.0 \\
...
\end{tabular}
\end{center}
\label{tab:rr_list}
\end{table}

\section*{Acknowledgments}

This work was performed at the Aspen Center for Physics, which is
supported by National Science Foundation grant PHY-1607611. We thank
Michael for an illuminating discussion. The research leading to these
results has received funding from the European Research Council under
the European Union's Seventh Framework Programme (FP/2007-2013) / ERC
Grant Agreement n. 308024. SK is partially supported by NSF grant
AST-1813881. NK is supported by NSF CAREER award 1455260.

This work presents results from the European Space Agency (ESA) space
mission Gaia. Gaia data are being processed by the Gaia Data
Processing and Analysis Consortium (DPAC). Funding for the DPAC is
provided by national institutions, in particular the institutions
participating in the Gaia MultiLateral Agreement (MLA). The Gaia
mission website is https://www.cosmos.esa.int/gaia. The Gaia archive
website is https://archives.esac.esa.int/gaia.

This project used public archival data from the Dark Energy Survey
(DES). Funding for the DES Projects has been provided by the
U.S. Department of Energy, the U.S. National Science Foundation, the
Ministry of Science and Education of Spain, the Science and Technology
Facilities Council of the United Kingdom, the Higher Education Funding
Council for England, the National Center for Supercomputing
Applications at the University of Illinois at Urbana-Champaign, the
Kavli Institute of Cosmological Physics at the University of Chicago,
the Center for Cosmology and Astro-Particle Physics at the Ohio State
University, the Mitchell Institute for Fundamental Physics and
Astronomy at Texas A\&M University, Financiadora de Estudos e
Projetos, Funda{\c c}{\~a}o Carlos Chagas Filho de Amparo {\`a}
Pesquisa do Estado do Rio de Janeiro, Conselho Nacional de
Desenvolvimento Cient{\'i}fico e Tecnol{\'o}gico and the
Minist{\'e}rio da Ci{\^e}ncia, Tecnologia e Inova{\c c}{\~a}o, the
Deutsche Forschungsgemeinschaft, and the Collaborating Institutions in
the Dark Energy Survey.  The Collaborating Institutions are Argonne
National Laboratory, the University of California at Santa Cruz, the
University of Cambridge, Centro de Investigaciones Energ{\'e}ticas,
Medioambientales y Tecnol{\'o}gicas-Madrid, the University of Chicago,
University College London, the DES-Brazil Consortium, the University
of Edinburgh, the Eidgen{\"o}ssische Technische Hochschule (ETH)
Z{\"u}rich, Fermi National Accelerator Laboratory, the University of
Illinois at Urbana-Champaign, the Institut de Ci{\`e}ncies de l'Espai
(IEEC/CSIC), the Institut de F{\'i}sica d'Altes Energies, Lawrence
Berkeley National Laboratory, the Ludwig-Maximilians Universit{\"a}t
M{\"u}nchen and the associated Excellence Cluster Universe, the
University of Michigan, the National Optical Astronomy Observatory,
the University of Nottingham, The Ohio State University, the OzDES
Membership Consortium, the University of Pennsylvania, the University
of Portsmouth, SLAC National Accelerator Laboratory, Stanford
University, the University of Sussex, and Texas A\&M University.
Based in part on observations at Cerro Tololo Inter-American
Observatory, National Optical Astronomy Observatory, which is operated
by the Association of Universities for Research in Astronomy (AURA)
under a cooperative agreement with the National Science Foundation.

The Legacy Surveys consist of three individual and complementary
projects: the Dark Energy Camera Legacy Survey (DECaLS; NOAO Proposal
ID \# 2014B-0404; PIs: David Schlegel and Arjun Dey), the
Beijing-Arizona Sky Survey (BASS; NOAO Proposal ID \# 2015A-0801; PIs:
Zhou Xu and Xiaohui Fan), and the Mayall z-band Legacy Survey (MzLS;
NOAO Proposal ID \# 2016A-0453; PI: Arjun Dey). DECaLS, BASS and MzLS
together include data obtained, respectively, at the Blanco telescope,
Cerro Tololo Inter-American Observatory, National Optical Astronomy
Observatory (NOAO); the Bok telescope, Steward Observatory, University
of Arizona; and the Mayall telescope, Kitt Peak National Observatory,
NOAO. The Legacy Surveys project is honored to be permitted to conduct
astronomical research on Iolkam Du'ag (Kitt Peak), a mountain with
particular significance to the Tohono O'odham Nation.

NOAO is operated by the Association of Universities for Research in
Astronomy (AURA) under a cooperative agreement with the National
Science Foundation.

The Legacy Survey team makes use of data products from the Near-Earth
Object Wide-field Infrared Survey Explorer (NEOWISE), which is a
project of the Jet Propulsion Laboratory/California Institute of
Technology. NEOWISE is funded by the National Aeronautics and Space
Administration.

The Legacy Surveys imaging of the DESI footprint is supported by the
Director, Office of Science, Office of High Energy Physics of the
U.S. Department of Energy under Contract No. DE-AC02-05CH1123, by the
National Energy Research Scientific Computing Center, a DOE Office of
Science User Facility under the same contract; and by the
U.S. National Science Foundation, Division of Astronomical Sciences
under Contract No. AST-0950945 to NOAO.

\appendix{

\section{RR Lyrae query}
\label{sec:rrl_query}
To select the RR Lyrae used in this paper we combined the information from two
Gaia classifiers, one is provided in the vari\_rrylae table and another one in the vari\_classifier\_result . We list the specific query below.

\begin{verbatim}
WITH x AS (
  SELECT vari_classifier_result.source_id
    FROM gaia_dr2.vari_classifier_result
    WHERE
vari_classifier_result.best_class_name::text
      ~~ 'RR%'::text
  UNION
  SELECT vari_rrlyrae.source_id
    FROM gaia_dr2.vari_rrlyrae
   ),
y AS (
  SELECT r.best_classification,
    r.solution_id,
    x.source_id, r.pf, r.pf_error,
    r.p1_o, r.p1_o_error,
    r.epoch_g,
    r.epoch_g_error,
    r.epoch_bp,
    r.epoch_bp_error,
    r.epoch_rp,
    r.epoch_rp_error,
    r.int_average_g,
    r.int_average_g_error,
    r.int_average_bp,
    r.int_average_bp_error,
    r.int_average_rp,
    r.int_average_rp_error,
    r.peak_to_peak_g,
    r.peak_to_peak_g_error,
    r.peak_to_peak_bp,
    r.peak_to_peak_bp_error,
    r.peak_to_peak_rp,
    r.peak_to_peak_rp_error,
    r.num_clean_epochs_g,
    r.num_clean_epochs_bp,
    r.num_clean_epochs_rp,
    FROM x
       LEFT JOIN gaia_dr2.vari_rrlyrae r
       USING (source_id)
    ),
z AS (
  SELECT y.*
    vr.classifier_name,
    vr.best_class_name,
    vr.best_class_score
    FROM y
    LEFT JOIN gaia_dr2.vari_classifier_result vr
    ON y.source_id = vr.source_id
    )
 SELECT z.best_classification,
  z.*, vv.*
  FROM z
  LEFT JOIN gaia_dr2.vari_time_series_statistics vv
  ON z.source_id = vv.source_id;

\end{verbatim}

\section{Transformation of equatorial coordinates to stream coordinates $\phione,\phitwo$}
\label{sec:rotation_matrix}

\begin{equation}
\begin{split}
\begin{pmatrix}
{\rm cos}(\phione)\,{\rm cos}(\phitwo)\\
{\rm sin}(\phione)\,{\rm cos}(\phitwo)\\
{\rm sin}(\phitwo) \\
\end{pmatrix}&=\\
\begin{pmatrix}
-0.44761231  &-0.08785756 & -0.88990128\\
-0.84246097  & 0.37511331 & 0.38671632\\
 0.29983786  & 0.92280606 & -0.2419219 \\
\end{pmatrix}\times&\\
\begin{pmatrix}{\rm cos}(\alpha)\,{\rm cos}(\delta)\\
{\rm sin}(\alpha)\,{\rm cos}(\delta)\\
{\rm sin}(\delta)\\
\end{pmatrix}&
\end{split}
\end{equation}

\section{The definition of splines used for the selection of RR Lyrae}

\begin{table}
\caption{The location of points defining the natural spline for the
  stream track shown on Fig.~\ref{fig:selection} to select RR Lyrae.}
\begin{center}

\begin{tabular}{cc}
\hline
$\phi_1$ & $\phi_2$ \\
deg & deg \\
\hline
-105.186 & 2.315 \\
-74.7184 & 2.148 \\
-54.4067 & 1.2797 \\
-25.4375 & -0.0562 \\
0.2019 & -0.8577 \\
25.8413 & -1.1917 \\
39.1605 & -1.9598 \\
66.9643 & 1.1127 \\
93.1032 & 3.2835 \\
108.0872 & 3.2167 \\
129.8974 & -0.1564 \\
162.5293 & -4.8653 \\
\hline
\end{tabular}
\label{tab:spline_track}
\end{center}
\end{table}
\begin{table}
\caption{The location of points defining the natural spline for the
  heliocentric distance track shown on Fig.~\ref{fig:selection} to
  select RR Lyrae.}
\begin{center}

\begin{tabular}{cc}
\hline
$\phi_1$ & Heliocentric distance \\
deg & kpc \\
\hline
-100.0 & 100.0 \\
-76.2169 & 58.0 \\
-69.9684 & 44.0 \\
-63.1307 & 39.441 \\
-44.2712 & 27.6252 \\
-25.0268 & 22.5353 \\
-5.3975 & 18.7178 \\
20.0052 & 15.9911 \\
42.3287 & 16.9 \\
61.958 & 18.5361 \\
90.0548 & 29.2612 \\
114.3027 & 44.8945 \\
\hline
\end{tabular}
\label{tab:dist_track}
\end{center}
\end{table}
\begin{table}
\caption{The location of points defining the spline for the
  track of proper motion along the stream  shown on Fig.~\ref{fig:selection} to select RR
  Lyrae.}
\begin{center}

\begin{tabular}{cc}
\hline
$\phi_1$ & $\mu_{\phi,1}$ \\
deg & mas\,yr$^{-1}$ \\
\hline
-77.4458 & 0.0408 \\
-49.626 & 1.3103 \\
-28.3432 & 2.3325 \\
-2.1956 & 3.1889 \\
24.2559 & 3.881 \\
44.0186 & 3.9108 \\
65.9096 & 2.6219 \\
88.4086 & 1.569 \\
103.6106 & 0.8984 \\
124.7415 & 0.0051 \\
\hline
\end{tabular}
\label{tab:pm_track}
\end{center}
\end{table}

\bibliography{references}

\label{lastpage}

\end{document}